# Permissible domain walls in monoclinic $M_{AB}$ ferroelectric phases

**Ido Biran[a] and Semën Gorfman[a]\***

[a]Department of Materials Science and Engineering, Tel Aviv University, Wolfson Building for Mechanical Engineering, Tel Aviv, 6997801, Israel

Correspondence email: gorfman@tauex.tau.ac.il

**Funding information**    Israel Science Foundation (grant No. 1561/18 to Semën Gorfman; grant No. 3455/21 to Semën Gorfman; grant No. 1365/23); United States - Israel Binational Science Foundation (award No. 2018161 to Semën Gorfman).

**Synopsis**    All the possibilities for permissible (mismatch-free) walls between monoclinic domains of pseudocubic ferroic perovskites are analyzed. Analytical expressions are derived for the orientation of such walls, the orientation relationship between the lattice vectors, and for the separation between Bragg peaks diffracted from matched domains.

**Abstract**    The concept of monoclinic ferroelectric phases has been extensively used over recent decades for the understanding of crystallographic structures of ferroelectric materials. Monoclinic phases have been actively invoked to describe the phase boundaries such as so-called Morphotropic Phase Boundary in functional perovskite oxides. These phases are believed to play a major role in the enhancement of such functional properties as dielectricity and electromechanical coupling through rotation of spontaneous polarization and / or modification of the rich domain microstructures. Unfortunately, such microstructures remain poorly understood due to the complexity of the subject. The goal of this work is to formulate the geometrical laws behind the monoclinic domain microstructures. Specifically, the result of our previous work (Gorfman *et al.*, 2022) is implemented to catalogue and outline some properties of permissible domain walls that connect "strain" domains with monoclinic ($M_A/M_B$ type) symmetry, occurring in ferroelectric perovskite oxides. The term "permissible" (Fousek & Janovec, 1969) pertains to the domain walls connecting a pair of "strain" domains without a lattice mismatch. It was found that 12 monoclinic domains may form pairs connected along 84 types of permissible domain walls. These contain 48 domain walls with fixed Miller indices (known as W-walls) and 36 domain walls whose Miller indices may change when free lattice parameters change as well (known as S-walls). We provide simple and intuitive analytical expressions that describe the orientation of these domain walls, the matrices of transformation between crystallographic basis vectors and, most importantly, the separation between Bragg peaks, diffracted from each of the 84 pairs of domains, connected along a permissible domain wall. It is shown that the orientation of domain wall may be described by the specific combination of the monoclinic distortion parameters $r = \frac{2}{\gamma - \alpha}\left(\frac{c}{a} - 1\right), f = \frac{\pi - 2\gamma}{\pi - 2\alpha}$ and $p = \frac{2}{\pi - \alpha - \gamma}\left(\frac{c}{a} - 1\right)$. The results of this work will enhance understanding and facilitate investigation (e.g., using single-crystal X-ray diffraction) of complex monoclinic domain microstructures in both crystals and thin films.

## 1. Introduction

Monoclinic ferroelectric phases (MFEP) have played an important role in understanding the structural mechanisms behind properties enhancement in functional ferroelectric materials, particularly in mixed-ion perovskite oxides. The concept of "monoclinic ferroelectrics" revolutionized the view on ferroelectricity by suggesting that spontaneous polarization can be rotated, rather than inverted or extended only (Davis *et al.*, 2007; Damjanovic, 2010). Evidence of MFEP was first reported by (Noheda *et al.*, 1999; Noheda *et al.*, 2000; Guo *et al.*, 2000) and supported by the splitting of Bragg reflections in high-resolution X-ray diffraction patterns. MFEPs were later incorporated into the higher-order Devonshire theory (Vanderbilt & Cohen, 2001) and invoked to explain the enhancement of the giant piezoelectric effect in PbZr$_{1-x}$Ti$_x$O$_3$ at the so-called Morphotropic Phase Boundary (MPB) (Fu & Cohen, 2000). The monoclinic space groups of ferroelectrics were used for many structural refinements based on X-ray and neutron scattering experiments (Gorfman & Thomas, 2010; Choe *et al.*, 2018; Zhang *et al.*, 2015; Zhang, Yokota *et al.*, 2014; Aksel *et al.*, 2011) or for the interpretation of the results of polarized light / birefringence experiments (Bokov *et al.*, 2010; Gorfman *et al.*, 2012). However, the true nature of the MFEP is still debated: it is not clear if the MFEP are truly long-range ordered or if the apparent long-range monoclinic order is "mimicked" by the so-called adaptive state, consisting of assemblies of locally tetragonal or rhombohedral nanodomains (Jin *et al.*, 2003; Viehland & Salje, 2014; Zhang, Xue *et al.*, 2014). Regardless of the true character of MFEP, the concept remains useful for the description of various phenomena in single crystal ferroelectrics (Noheda *et al.*, 2001; Choe *et al.*, 2018; Gorfman *et al.*, 2012), ferro/piezoceramics (Liu *et al.*, 2017; Zhang, Yokota *et al.*, 2014), epitaxial thin films (Wang *et al.*, 2003; von Helden *et al.*, 2018; Braun *et al.*, 2018; Schmidbauer *et al.*, 2017; de Oliveira Guimarães *et al.*, 2022)  and shape memory alloys (Bhattacharya, 2003).

Besides the interesting intrinsic properties of MFEP, rich microstructures of monoclinic domains (MDs) and domain walls (DWs) between them attract a great deal of  interest (Nakajima *et al.*, 2022; Mantri & Daniels, 2021). Any domain microstructures may underpin exotic physical properties such as giant electromechanical coupling (Hu *et al.*, 2020), enhanced dielectric permittivity (Trolier-McKinstry *et al.*, 2018), superelasticity (Viehland & Salje, 2014), shape memory effect (Bhattacharya, 2003) and domain-wall superconductivity (Catalan *et al.*, 2012). These microstructure-driven properties are particularly diverse when individual domains host several order parameters (e.g., electrics, magnetic and elastic). Remarkably, such properties are often absent in a single domain. Their appearance and magnitude depend on the mobility of DWs. MFEPs should have rich and volatile domains microstructures. Therefore, the properties of DWs in MFEPs (such as crystallographic orientation and mobility) are relevant for the understanding of physical properties of materials. Although the algorithms for the prediction of DWs between domains of different symmetry are known (Fousek & Janovec, 1969; Sapriel, 1975; Authier, 2003), the underlying complexity of the subjects prevents any comprehensive understanding of domains microstructures of MFEPs.

The aim of this work is to describe the geometry of permissible DWs (PDWs) between domains of MFEPs. The term permissible (coined by Fousek and Janovec (Fousek & Janovec, 1969), see also (Sapriel, 1975)) denotes a planar DW connecting two domains without any lattice mismatch. For example, tetragonal domains are permitted to connect along DWs of six different orientations (with the Miller indices belonging to the family {110}), rhombohedral domains are permitted to connect along DWs of 12 different orientations [with the Miller indices belonging to the families {110} and {100} and exhibiting different physical properties (such as e.g. scattering of light) (Qiu *et al.*, 2020)].

We demonstrate that, most generally, MDs are permitted to connect along 84 types of DWs of 45 different orientations and five different orientational families. More specifically, we show that all the 84 DW contain 48 prominent DWs (*W*-walls) which have fixed crystallographic orientation and 36 *S-walls* which change their orientation when free monoclinic lattice parameters change too. In addition, we present the analytical expressions for the matrices of transformation between the lattice basis vectors of matched domains and for the separation between Bragg peaks, diffracted from such domains. The presented equations create the direct path for the calculations of DW-related quantities, such as angles between polarization directions, the direction of DW-motion under electric field and so on.

## 2. Monoclinic ferroelectric phases: important definitions

This paper implements the list of notations and abbreviations introduced by Gorfman *et al.* (2022). Appendix A summarizes the most important ones. This section describes the definitions, relevant for the description of the monoclinic phases of ferroelectric perovskites.

According to Fu & Cohen (2000), MFEPs can be of $M_A/M_B$ or alternatively $M_C$ types. These types differ from one another by the set of independent pseudocubic lattice parameters and by the direction in which spontaneous polarization may develop. Note that while the spontaneous polarization vector, **P** (SP), is not mandatory in ferroelastic domains, it exists in practice for the case of ferroic perovskite oxides. Even if the magnitude of such polarization is zero, it is still useful to consider the potential SP direction(s) for domain referencing and numbering.

This paper focuses on $M_A/M_B$ case. The case of $M_C$ domains is described in a follow-up paper.

## 2.1. The definition of M_A/M_B monoclinic domains

The crystallographic structures of the $M_A/M_B$ phases of perovskite oxides belong to the space group types $Cm$, $Cc$ (Zhang, Yokota *et al.*, 2014). These structures are obtained by the symmetry-lowering phase transitions from those described by the rhombohedral ($R$) space group types $R3m$, $R3c$. The mirror ($m$) / glide ($c$) plane is parallel to two mutually perpendicular face-diagonal and the edge of the pseudocubic unit cell. These space group types allow for rotation of the polar axis (e.g., the direction of spontaneous polarization vector) within this mirror plane. Additionally, these space groups permit any distortion of the unit cell that maintains the mirror plane. Both the distortion of the pseudocubic unit cell (alongside the mirror plane) and the polar axis direction are shown in Figure 1a.

### 2.1.1. The numeration of the monoclinic domains and the potential spontaneous polarization

It is convenient to illustrate the monoclinic domains using stereographic projection and the corresponding potential spontaneous polarization direction (SPD). Since $M_A/M_B$ domains arise from the transition from the rhombohedral $R$ phase, we define the SPD by a small rotation angle $\rho$ from any of the four body-diagonal directions $\langle 111 \rangle$ towards any of the three adjacent unit cell edges $\langle 001 \rangle$. We mark the corresponding 12 monoclinic domains as $M_{nm}$ where the first index, $n$, lists the SPDs $\boldsymbol{R}_n$ in the "parent" rhombohedral domain. In this case, $\boldsymbol{R}_1 = [111]$, $\boldsymbol{R}_2 = [\bar{1}11]$, $\boldsymbol{R}_3 = [1\bar{1}1]$ and $\boldsymbol{R}_4 = [\bar{1}\bar{1}1]$. The second index, $m$ ($m = 1 ... 3$), marks the pseudocubic axis $\boldsymbol{T}_m$ so that $\boldsymbol{T}_1 = \pm[100]$, $\boldsymbol{T}_2 = \pm[010]$, $\boldsymbol{T}_3 = \pm[001]$ to which the polarization rotates. For example, the monoclinic domain $M_{13}$ has its SPD rotated from $[111]$ towards $[001]$, while $M_{21}$ has its SPD rotated from $[\bar{1}11]$ towards $[\bar{1}00]$. The SPDs in all 12 monoclinic domains are shown on the stereographic projections in Figure 1b.

In the following, we express the coordinates of the SPD relative to the axes of the Cartesian coordinate system, that are nearly parallel to the pseudocubic basis vectors. For the cases of domains $M_{13}$ we obtain

$$[\boldsymbol{P}]_{13} = [11x] \tag{1}$$

Here, we introduced the notation:

$$x = \sqrt{2}\cot(\rho_0 - \rho), \tag{2}$$

with $\rho_0 \approx 54.7°$ the angle between the body diagonal and the edge of a cube, so that $\cos\rho_0 = \frac{1}{\sqrt{3}}$, $\sin\rho_0 = \frac{\sqrt{2}}{\sqrt{3}}$. Assuming the SPD rotation angle $\rho$ is small and keeping the first term in the Taylor expansion with respect to $\rho$, we can rewrite equation (2) as

$$x = 1 + \frac{3}{\sqrt{2}}\rho + O(\rho^2) \tag{3}$$

Note that the cases of $\rho > 0$ and $\rho < 0$ are referred to as $M_A$ and $M_B$ phases correspondingly.

### 2.1.2. Pseudocubic lattice parameters of the monoclinic M_A/M_B domains

Figure 1a shows the $M_A/M_B$ distortion of the pseudocubic unit cell. The corresponding pseudocubic lattice parameters $a_i, \alpha_i$ $(i = 1 \dots 3)$ are described by four independent variables: $a, c, \alpha, \gamma$ (Gorfman & Thomas, 2010; Aksel *et al.*, 2011; Choe *et al.*, 2018): e.g. for the domain $M_{13}$ $a_1 = a_2 = a, a_3 = c, \alpha_1 = \alpha_2 = \alpha, \alpha_3 = \gamma$. The corresponding matrix of dot product is

$$[G]_{13} = \begin{bmatrix} a^2 & a^2 \cos\gamma & ac\cos\alpha \\ a^2\cos\gamma & a^2 & ac\cos\alpha \\ ac\cos\alpha & ac\cos\alpha & c^2 \end{bmatrix} = a^2([I] + [G']_{13}) \tag{4}$$

Here, $[I]$ is the unitary matrix and

$$[G']_{13} = \begin{bmatrix} 0 & B & A \\ B & 0 & A \\ A & A & C \end{bmatrix} \tag{5}$$

Assuming that the monoclinic distortion is small and keeping the first power of $\left(\frac{c}{a} - 1\right), \frac{\pi}{2} - \alpha, \frac{\pi}{2} - \gamma$ we can write:

$$\begin{aligned} C &\approx 2\left(\frac{c}{a} - 1\right) \\ A &\approx \left(\frac{\pi}{2} - \alpha\right) = \Delta\alpha \\ B &\approx \left(\frac{\pi}{2} - \gamma\right) = \Delta\gamma \end{aligned} \tag{6}$$

The resulting **monoclinic crystal lattice** is invariant with respect to $N_M = 4$ symmetry operations of the holohedry point group $\frac{2}{m}$. The parent **cubic crystal lattice** is invariant with respect to $N_C = 48$ operations of the holohedry point group $m3m$. Because the monoclinic distortion may commence from any of these 48 equivalent variants, there are $\frac{N_C}{N_M} = 12$ variants of the monoclinic domain's variants. These are listed in the Table 1, which contains domain identifications, $M_{nm}$, the $[G']$ metric tensors, the SPD, and the lattice parameters $a_1, a_2, a_3, \alpha_1, \alpha_2, \alpha_3$.

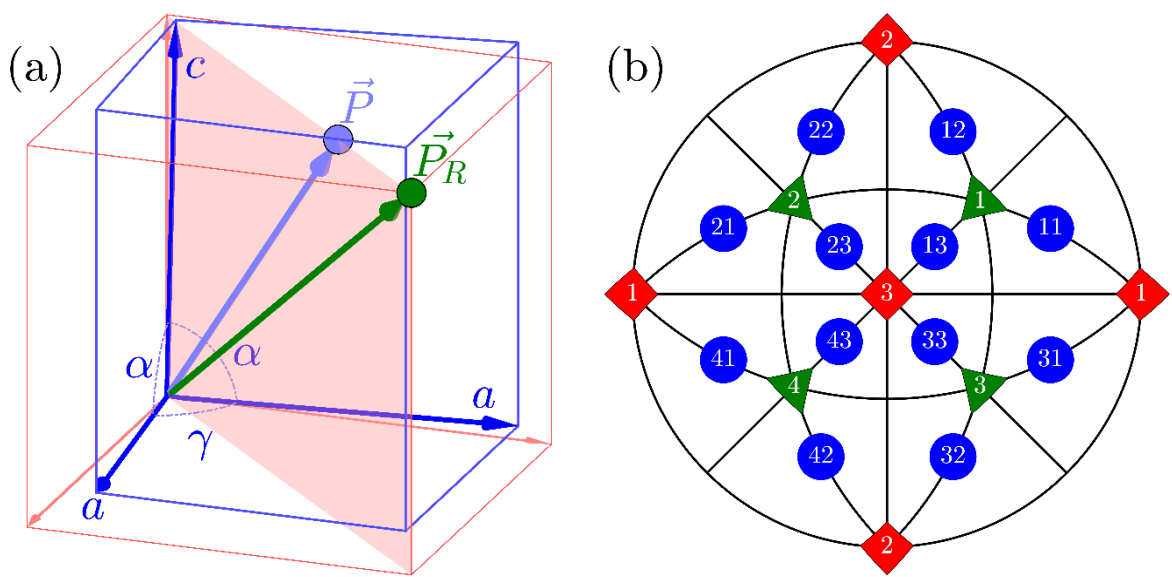

**Figure 1** Schematic illustration of the $M_A/M_B$ monoclinic domains and numeration of their variants. (a) The unit cell distortion along with the rotation of the spontaneous polarization direction (SPD) (if such polarization is present). (b) The stereographic projection, showing these directions for the domains of tetragonal (red squares), rhombohedral (green triangles) and monoclinic (blue circles) symmetry. The tetragonal domains (1), (2), (3) correspond to the SPD along [100], [010] and [001] respectively. The rhombohedral domains (1), (2), (3) and (4) correspond to the SPD along [111], [$\bar{1}$11], [1$\bar{1}$1] and [11$\bar{1}$] directions, respectively. The SPDs within the 12 monoclinic domains are further explained in Table 1.

**Table 1** The definition of the 12 monoclinic ($M_A/M_B$ type) domains variants. The first column contains the domain variant identifier (as also displayed in Figure 1b). The second column contains the twinning matrix (the definition of this matrix is explained by Gorfman *et al.* (2022) but also presented in equation (A4)). The third column contains the SPD for each domain, relative to the domain-related crystallographic coordinate system. The fourth column contains the pseudocubic lattice parameters expressed in terms of free parameters $a, c, \alpha, \gamma$. The notations $\breve{\alpha} = \pi - \alpha$ and $\breve{\gamma} = \pi - \gamma$ are used. The last column contains the reduced matrix $[G']_{mn} = \frac{[G]_{mn}}{a^2} - [I]$. The calculations of the $[G']_{mn}$ and corresponding lattice parameters are done using equation (A5).

| Domain "name" | Twinning matrix $[T]$ | $[\boldsymbol{P}]_{mn}$ | Pseudocubic LP | $[G']_{mn}$ |
|---|---|---|---|---|
| $M_{11}$ | $\begin{bmatrix} 0 & 1 & 0 \\ 0 & 0 & 1 \\ 1 & 0 & 0 \end{bmatrix}$ | $[x11]$ | $c\ a\ a\quad \gamma\ \alpha\ \alpha$ | $\begin{bmatrix} C & A & A \\ A & 0 & B \\ A & B & 0 \end{bmatrix}$ |
| $M_{12}$ | $\begin{bmatrix} 0 & 0 & 1 \\ 1 & 0 & 0 \\ 0 & 1 & 0 \end{bmatrix}$ | $[1x1]$ | $a\ c\ a\quad \alpha\ \gamma\ \alpha$ | $\begin{bmatrix} 0 & A & B \\ A & C & A \\ B & A & 0 \end{bmatrix}$ |
| $M_{13}$ | $\begin{bmatrix} 1 & 0 & 0 \\ 0 & 1 & 0 \\ 0 & 0 & 1 \end{bmatrix}$ | $[11x]$ | $a\ a\ c\quad \alpha\ \alpha\ \gamma$ | $\begin{bmatrix} 0 & B & A \\ B & 0 & A \\ A & A & C \end{bmatrix}$ |
| $M_{21}$ | $\begin{bmatrix} 0 & 1 & 0 \\ 0 & 0 & 1 \\ \bar{1} & 0 & 0 \end{bmatrix}$ | $[\bar{x}11]$ | $c\ a\ a\quad \gamma\ \breve{\alpha}\ \breve{\alpha}$ | $\begin{bmatrix} C & \bar{A} & \bar{A} \\ \bar{A} & 0 & B \\ \bar{A} & B & 0 \end{bmatrix}$ |
| $M_{22}$ | $\begin{bmatrix} 0 & 0 & 1 \\ \bar{1} & 0 & 0 \\ 0 & 1 & 0 \end{bmatrix}$ | $[\bar{1}x1]$ | $a\ c\ a\quad \alpha\ \breve{\gamma}\ \breve{\alpha}$ | $\begin{bmatrix} 0 & \bar{A} & \bar{B} \\ \bar{A} & C & A \\ \bar{B} & A & 0 \end{bmatrix}$ |
| $M_{23}$ | $\begin{bmatrix} \bar{1} & 0 & 0 \\ 0 & 1 & 0 \\ 0 & 0 & 1 \end{bmatrix}$ | $[\bar{1}1x]$ | $a\ a\ c\ \ \alpha\ \breve{\alpha}\ \bar{\gamma}$ | $\begin{bmatrix} 0 & \bar{B} & \bar{A} \\ \bar{B} & 0 & A \\ \bar{A} & A & C \end{bmatrix}$ |
| $M_{31}$ | $\begin{bmatrix} 0 & \bar{1} & 0 \\ 0 & 0 & 1 \\ 1 & 0 & 0 \end{bmatrix}$ | $[x\bar{1}1]$ | $c\ a\ a\ \breve{\gamma}\ \alpha\ \breve{\alpha}$ | $\begin{bmatrix} C & \bar{A} & A \\ \bar{A} & 0 & \bar{B} \\ A & \bar{B} & 0 \end{bmatrix}$ |
| $M_{32}$ | $\begin{bmatrix} 0 & 0 & 1 \\ 1 & 0 & 0 \\ 0 & \bar{1} & 0 \end{bmatrix}$ | $[1\bar{x}1]$ | $a\ c\ a\ \breve{\alpha}\ \gamma\ \breve{\alpha}$ | $\begin{bmatrix} 0 & \bar{A} & B \\ \bar{A} & C & \bar{A} \\ B & \bar{A} & 0 \end{bmatrix}$ |
| $M_{33}$ | $\begin{bmatrix} 1 & 0 & 0 \\ 0 & 1 & 0 \\ 0 & 0 & \bar{1} \end{bmatrix}$ | $[1\bar{1}x]$ | $a\ a\ c\ \breve{\alpha}\ \alpha\ \bar{\gamma}$ | $\begin{bmatrix} 0 & \bar{B} & A \\ \bar{B} & 0 & \bar{A} \\ A & \bar{A} & C \end{bmatrix}$ |
| $M_{41}$ | $\begin{bmatrix} 0 & \bar{1} & 0 \\ 0 & 0 & 1 \\ \bar{1} & 0 & 0 \end{bmatrix}$ | $[\bar{x}\bar{1}1]$ | $c\ a\ a\ \breve{\gamma}\ \breve{\alpha}\ \alpha$ | $\begin{bmatrix} C & A & \bar{A} \\ A & 0 & \bar{B} \\ \bar{A} & \bar{B} & 0 \end{bmatrix}$ |
| $M_{42}$ | $\begin{bmatrix} 0 & 0 & 1 \\ \bar{1} & 0 & 0 \\ 0 & \bar{1} & 0 \end{bmatrix}$ | $[\bar{1}\bar{x}1]$ | $a\ c\ a\ \breve{\alpha}\ \bar{\gamma}\ \alpha$ | $\begin{bmatrix} 0 & A & \bar{B} \\ A & C & \bar{A} \\ \bar{B} & \bar{A} & 0 \end{bmatrix}$ |
| $M_{43}$ | $\begin{bmatrix} \bar{1} & 0 & 0 \\ 0 & 1 & 0 \\ 0 & 0 & \bar{1} \end{bmatrix}$ | $[\bar{1}\bar{1}x]$ | $a\ a\ c\ \breve{\alpha}\ \breve{\alpha}\ \gamma$ | $\begin{bmatrix} 0 & B & \bar{A} \\ B & 0 & \bar{A} \\ \bar{A} & \bar{A} & C \end{bmatrix}$ |

## 2.2. Domain pairs

12 ferroelastic domains (Table 1) can form 66 domains pairs. Some of these pairs can be connected via PDWs and some of them cannot. Before analyzing PDWs between various pairs of monoclinic domains, we will introduce five different pair types. These types are referred to as "*R-sibling*", "*R-planar*", "*R-semi-planar*", "*R-semi-crossed*" and "*R-crossed*". Each type has its own angle between the SPDs and its own expressions for the indices of PDWs. Accordingly, we expect different properties from various domain pair types, with respect to e.g., DW motion under an external electric field. Table 2 presents the information about all five domain pairs, including pair name, abbreviation, formal definition, the angles between SPDs and the reference figure.

### 2.2.1. Domain pairs of the type "*R-sibling*"

We will use the term "*R-sibling*" for 12 pairs of monoclinic domains $M_{nk}$ $M_{nl}$ such that the members of each pair originate from the same parent / rhombohedral domain $R_n$. Three *R-sibling* pairs can be formed for each $R_n$: $M_{n1}$ $M_{n2}$, $M_{n2}$ $M_{n3}$ and $M_{n3}$ $M_{n1}$. All such pairs are illustrated on the stereographic projections (viewed along [001] and [110] directions) in Figure 2.

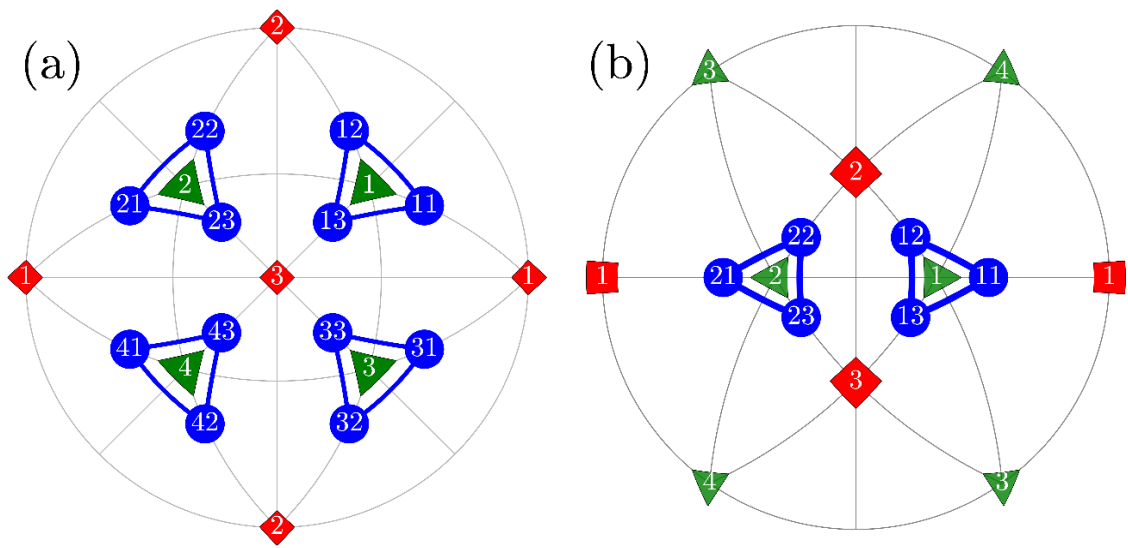

**Figure 2** Schematic illustration of the "*R-sibling*" type of monoclinic domain pairs. The term "*R-sibling*" refers to the case when both pair members originate from the same *R* domain. The figure includes: (a) Stereographic projection viewed along [001] direction, showing the SPDs in the 12 monoclinic domains. (b) Stereographic direction viewed along the direction [110] highlighting the *sibling* pair types, originating from $R_1$ and $R_2$.

### 2.2.2. Domain pairs of the type "*R-planar*"

We will use the term **"*R-planar*"** for six pairs of monoclinic domains $M_{mk}M_{nk}$, originating from different rhombohedral domains $R_m$ and $R_n$ ($m \neq n$) but such that $k = \mathcal{L}(m,n)$, where $\mathcal{L}(m,n)$ marks the pseudocubic axis that is parallel to the $\boldsymbol{R_m R_n}$ *plane* so that:

$$\mathcal{L}(1,2) = 1 \quad \mathcal{L}(1,3) = 2 \quad \mathcal{L}(1,4) = 3 \quad \mathcal{L}(2,3) = 3 \quad \mathcal{L}(2,4) = 2 \quad \mathcal{L}(3,4) = 1 \quad (7)$$

All the *R-planar* domain pairs are illustrated in Figure 3 on the same type of stereographic projection as in Figure 2.

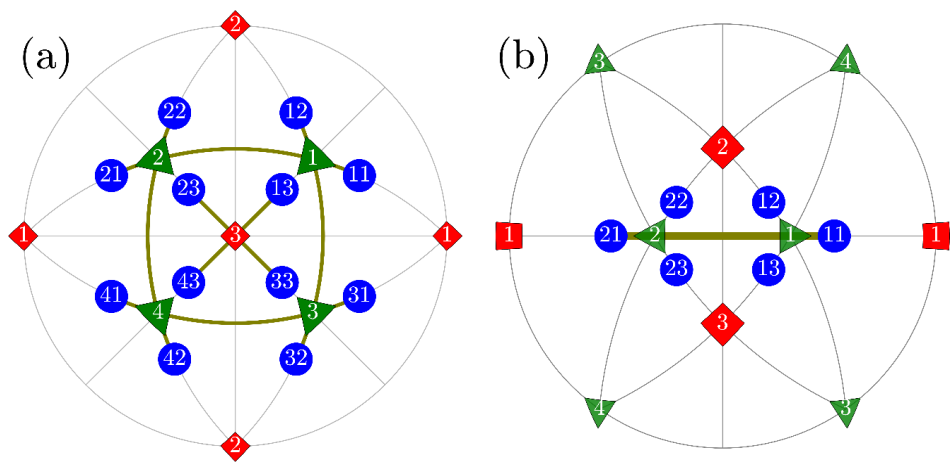

**Figure 3** The same as Figure 2 but for the case of "*R-planar*" type of DW pairs.

### 2.2.3. Domain pairs of the type "*R-semi-planar*"

We use the term **"*R-semi-planar*"** for 12 pairs of monoclinic domains $M_{nk}M_{mk}$ originating from different rhombohedral domains $R_m$ and $R_n$ $(m \neq n)$ but such that $k \neq \mathcal{L}(m,n)$. Each $R_mR_n$ pair produces two monoclinic domains pairs of this type, e.g., $M_{12}\,M_{22}$ and $M_{13}\,M_{23}$ for the case of $R_1R_2$. All the *R-semi-planar* domain pairs are illustrated in Figure 4.

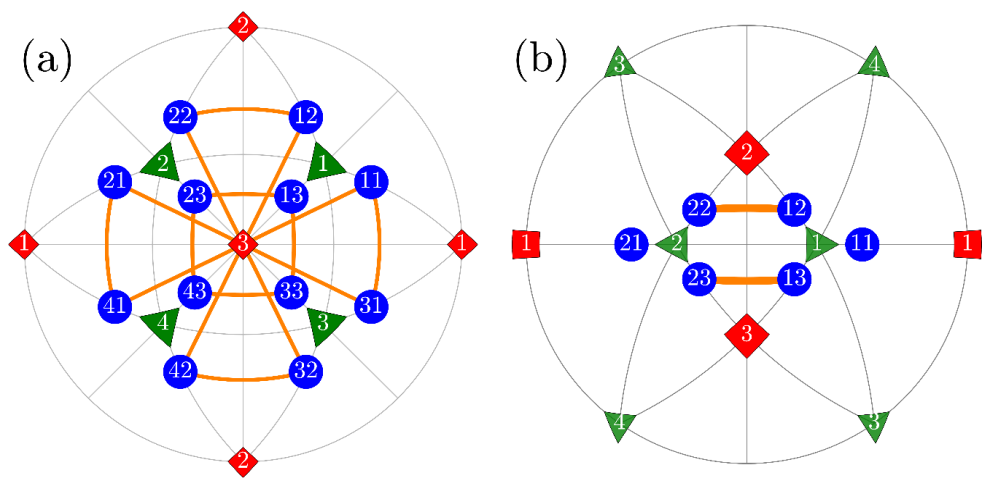

**Figure 4** The same as Figure 2 but for the case of "R-semi-planar" type of DW pair.

### 2.2.4. Domain walls of the type "*R-semi-crossed*"

We will use the term "*R-semi-crossed*" for 12 pairs of monoclinic domains $M_{mk}$ and $M_{nl}$, originating from different rhombohedral domains $R_m$ and $R_n$ $(m \neq n)$ and such that both $k \neq \mathcal{L}(m,n)$ and $l \neq \mathcal{L}(m,n)$. In addition, $k \neq l$ because the cases of $k = l$ are already included in the "R-semi-planar" types of domain pairs. Each $R_mR_n$ pair produces two pairs of monoclinic domains of this type e.g.

$M_{12}M_{23}$ and $M_{13}M_{22}$ for the case of $R_1R_2$. All the *R-semi-crossed* pairs of domains are illustrated in Figure 5.

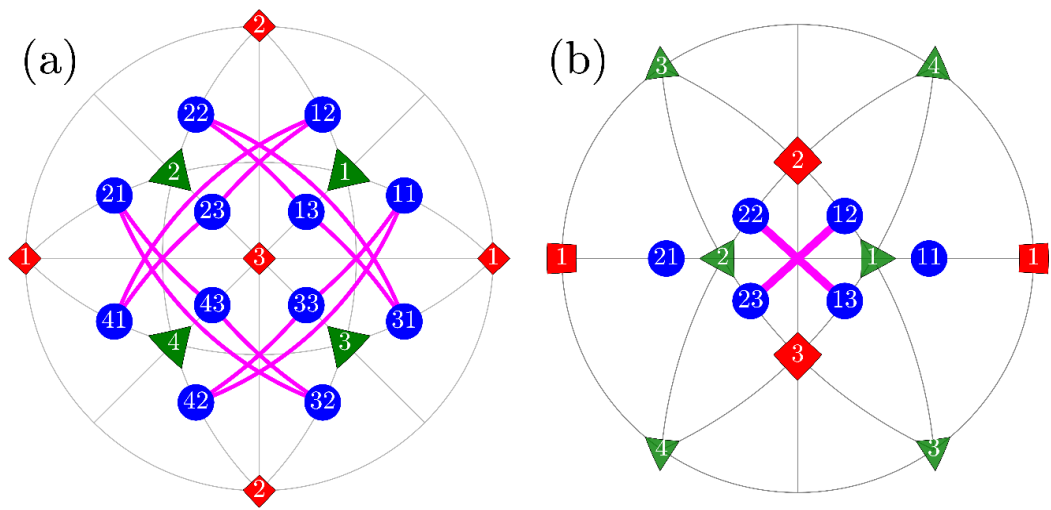

**Figure 5** The same as the Figure 2 bit for the case of *R-semi-crossed* twin domain pairs.

### 2.2.5 Domain pairs of the type "*R-crossed*".

We will finally use the term "***R-crossed***" for 24 pairs of monoclinic domains $M_{mk}$ and $M_{nl}$ such that $m \neq n, k \neq l$ while either $k = \mathcal{L}$ or $l = \mathcal{L}$. Each $R_mR_n$ pair produces four pairs of monoclinic domains of this type, for example $M_{11}M_{23}, M_{11}M_{22}, M_{13}M_{21}$ and $M_{12}M_{21}$ for the case of $R_1R_2$. All the *R-crossed* domains are illustrated in Figure 6. We will see later that these pair types may generally not be connected via PDWs.

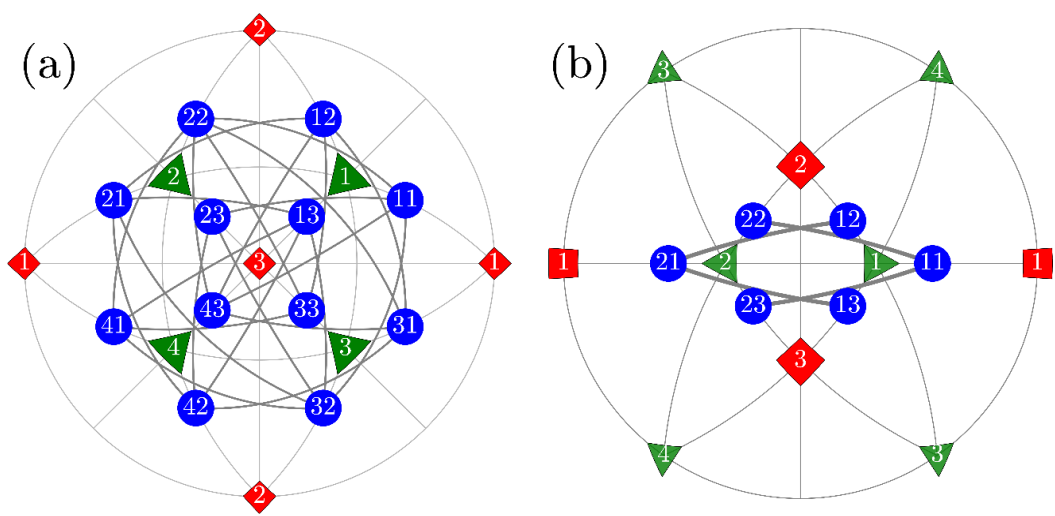

**Figure 6** The same as Figure 2 but for the case of *R-crossed* type of domain pair.

**Table 2**  The definitions of monoclinic $M_A/M_B$ domains pair types. The first two columns contain the domain pair name (full and short), the third column defines the pair, the fourth column lists the angle $\xi$ between SPDs as a function of $\rho$. This angle can be calculated by using Table 1, equation (2) and keeping the first power of $\rho$ in the Taylor series expansion. The fifth column contains the number of the corresponding domain pairs, the last column refers to the corresponding figure.

| Full name | Short name | Formal definition | $\xi, \bar{\xi}$ | Number of pairs | Figure |
|---|---|---|---|---|---|
| *R-sibling* | RSB | $M_{nk}\,M_{nl}$ | $\sqrt{3}\rho$ | 12 | Figure 2 |
| *R-planar* | RP | $M_{mk}M_{nk},$ $k = \mathcal{L}(m,n)$ | $\xi_R + 2\rho$ | 6 | Figure 3 |
| *R-semi-planar* | RSP | $M_{mk}M_{nk},$ $k \neq \mathcal{L}(m,n)$ | $\xi_R - \rho$ | 12 | Figure 4 |
| *R-semi-crossed* | RSC | $M_{mk}M_{nl}$ $k \neq \mathcal{L}(m,n), l \neq \mathcal{L}(m,n)$ $k \neq l$ | $\xi_R - \rho$ | 12 | Figure 5 |
| *R-crossed* | RC | $M_{mk}M_{nl}$ $k \neq \mathcal{L}(m,n), l = \mathcal{L}(m,n)$ Or $k = \mathcal{L}(m,n), l \neq \mathcal{L}(m,n)$ | $\xi_R - \dfrac{\rho}{2}$ | 24 | Figure 6 |

## 3. The orientation of PDWs between different pairs of domains

According to Fousek & Janovec (1969), the term PDW stands for a planar DW, that enables mismatch-free connection of one domain to another. PDWs are parallel to lattice planes with specific Miller indices ($hkl$) which have the same two-dimensional lattice parameters in both domains connected. For any two arbitrary domains, described by the matrices of dot products $[G]_n$ and $[G]_m$, such plane should satisfy the equations $hx_1 + kx_2 + lx_3 = 0$ and $\Delta G_{ij} x_i x_j = 0$ (here $[\Delta G] = [G]_n - [G]_m$). The key steps (see Gorfman *et al.* (2022)) for finding the orientation of the PDWs between two arbitrary domains (Table 1) are:

- Finding the eigenvalues ($\lambda_1, \lambda_2$ and $\lambda_3$) of $[\Delta G]$ or, equivalently, $[\Delta G'] = [G']_n - [G']_m$.

- Checking if these domains have PDWs. This is the case if at least one eigenvalue is zero (e.g., $\lambda_2 = 0$). This condition is fulfilled if and only if $|\Delta G'| = 0$.

- Rearranging the eigenvalues so that $\lambda_2 = 0$, $\lambda_3 > 0$. Importantly, for all the cases considered in this paper $\Delta G_{11} + \Delta G_{22} + \Delta G_{33} = 0$, which means that $\lambda_1 + \lambda_2 + \lambda_3 = 0$ and $\lambda_1 = -\lambda_3$.

- Forming orthogonal matrix $[V]$ ( $[V]^{-1} = [V]^T$ ) whose columns are the corresponding normalized eigenvectors of $[\Delta G']$.

- Finding the PDW indices (the coordinates of the PDW normal with respect to the reciprocal basis vector $\boldsymbol{a}^*_{im}$) according to $h'_i = V_{i1} - V_{i3}$ or $h'_i = V_{i1} + V_{i3}$ .

- When possible, $h'_i$ can be extended to the nearest all-integer values to get the Miller indices of the corresponding DW $hkl$.

Besides the ability to calculate the Miller indices of the PDW, this approach provides the basis for the calculation of orientation relationship between domain's basis vectors and separation of Bragg peaks, diffracted from matched pair of domains. This possibility is the main advantage of this approach over those already existing (Fousek & Janovec, 1969). The relevant information for calculating these quantities is given further in section 5.

### 3.1. Permissible domains walls connecting domain pairs of the type "*R-sibling*".

We will demonstrate the derivation of the PDWs connecting the representative domain pair $M_{12}$ $M_{13}$ and obtain the similar results for all the other pairs of this type analogously. Using the last column of Table 1 and equations (6) we obtain

$$[G']_{13} - [G']_{12} = (B - A)[\Delta G_{RSB}] \tag{8}$$

The following notation was introduced here:

$$[\Delta G_{RSB}] = \begin{pmatrix} 0 & 1 & \bar{1} \\ 1 & r & 0 \\ \bar{1} & 0 & \bar{r} \end{pmatrix} \tag{9}$$

and

$$r = \frac{C}{A - B} \approx \frac{2}{\gamma - \alpha}\left(\frac{c}{a} - 1\right) \tag{10}$$

The eigenvalues of the $[\Delta G_{RSB}]$ can be found trivially as $-\lambda_{RSB}, 0, \lambda_{RSB}$ with

$$\lambda_{RSB} = \sqrt{2 + r^2} \tag{11}$$

The corresponding eigenvalues of the matrix $[G']_{13} - [G']_{12}$ are $-\lambda_{3\,RSB}, 0, \lambda_{3\,RSB}$ with

$$\lambda_{3\,RSB} = (B - A)\sqrt{2 + r^2} \approx (\alpha - \gamma)\sqrt{2 + r^2} \tag{12}$$

The orthogonal matrix of eigenvectors of $[\Delta G_{RSB}]$ (as well as $[G']_{13} - [G']_{12}$) can be expressed as

$$[V_{RSB}] = \frac{1}{2\lambda_{RSB}}\begin{pmatrix} 2 & 2r & 2 \\ r - \lambda_{RSB} & 2 & r + \lambda_{RSB} \\ r + \lambda_{RSB} & 2 & r - \lambda_{RSB} \end{pmatrix} \tag{13}$$

Accordingly, two PDWs normal to the vectors $RSB_i^{(1,2)} \sim (V_{i1} \mp V_{i3})$ are possible.

$$[RSB^{(1)}] = \begin{pmatrix} 0 \\ \bar{1} \\ 1 \end{pmatrix}$$

$$[RSB^{(2)}] = \begin{pmatrix} 2 \\ r \\ r \end{pmatrix} \tag{14}$$

The $[RSB^{(1)}]$ normal has fixed coordinates that do not depend on the free lattice parameters. According to Fousek & Janovec (1969), such a wall can therefore be referred to as *W-wall*. On the contrary, the $[RSB^{(2)}]$ depends on the monoclinic distortion parameter $r$ and according to Fousek & Janovec (1969) it can be referred to as *S-wall* (*"strange"* DW). Although the monoclinic distortion parameters $C, A, B$ are small, the value of $r$ (as a ratio of $C$ and $A - B$) is not. This means that even a small change of monoclinic distortion may cause significant reorientation of the PDW. Table 3 highlights several favorable cases of the monoclinic distortion parameter $r$ which sets the $S$- wall to have rational "Miller" indices. For example, $r = 2$ (when $1 - \frac{c}{a} = \gamma - \alpha$) creates PDW along (111) plane. Approaching $r = \infty$ (e.g., $\alpha = \gamma$) would mean the appearance of PDW parallel to (011).

**Table 3**  The special cases of monoclinic distortion, leading to the appearance of S-walls with rational "Miller indices". The first column contains the relevant condition for the lattice parameters, the second column contains the corresponding value of $r$. The third column contains the eigenvalue $\lambda_{3\,RSB}$ of the matrix $[G']_{13} - [G']_{12}$. The condition of mismatch-free connection is only relevant for the case if $\lambda_{3\,RSB} \neq 0$ (otherwise the domains may connect along any plane). The last column contains the Miller indices of the DW.

| Lattice parameters | $r$ | $\lambda_{3\,RSB}$ | S-wall orientation |
|---|---|---|---|
| $c = a$ | 0 | $(\alpha - \gamma)\sqrt{2}$ | (100) |

| $\alpha = \gamma$ | $\infty$ | $2(\frac{c}{a} - 1)$ | (011) |
| $\gamma - \alpha = \left(1 - \frac{c}{a}\right)$ | 2 | $+\sqrt{6}(\alpha - \gamma)$ | (111) |

### 3.2.   PDWs connecting domain pairs of the type "*R-planar*".

We will demonstrate the derivation of the PDWs connecting the representative domain pair $M_{11}$ $M_{21}$ and obtain the similar results for the other pairs of this type analogously. Using the last column of the Table 1 and equations (6)

$$[G']_{21} - [G']_{11} = -2A[\Delta G_{RP}] \approx -2\Delta\alpha[\Delta G_{RP}] \tag{15}$$

Here, we introduced the following notation.

$$[\Delta G_{RP}] = \begin{pmatrix} 0 & 1 & 1 \\ 1 & 0 & 0 \\ 1 & 0 & 0 \end{pmatrix} \tag{16}$$

It is straightforward to see that the eigenvalues of $[\Delta G_{RP}]$ are $-\sqrt{2}, 0, \sqrt{2}$. Similarly, the eigenvalues of the matrix $[G']_{21} - [G']_{11}$ are $-\lambda_{3\,RP}, 0, \lambda_{3\,RP}$ with

$$\lambda_{3\,RP} = -2\sqrt{2}\Delta\alpha \tag{17}$$

The orthogonal matrix of eigenvectors of both $[\Delta G_{RP}]$ and $[G']_{21} - [G']_{11}$ is

$$[V_{RP}] = \frac{1}{2}\begin{pmatrix} -\sqrt{2} & 0 & \sqrt{2} \\ 1 & -\sqrt{2} & 1 \\ 1 & +\sqrt{2} & 1 \end{pmatrix} \tag{18}$$

Accordingly, two PDWs normal to the vectors $RP_i^{(1,2)} \sim (V_{RP\,i1} \mp V_{RP\,i3})$ exist.

$$[RP^{(1)}] = \begin{pmatrix} 1 \\ 0 \\ 0 \end{pmatrix}$$

$$[RP^{(2)}] = \begin{pmatrix} 0 \\ 1 \\ 1 \end{pmatrix} \tag{19}$$

Both are *W-walls*, i.e., the crystallographic orientation of these walls does not depend on the values of the lattice parameters.

### 3.3. PDWs connecting domain pairs of the type "*R-semi-planar*".

We will demonstrate the derivation of the PDWs connecting the representative domain pair $M_{12}$ $M_{22}$ and obtain the similar results for the other pairs of this type analogously. According to the last column of Table 1 and equations (6):

$$[G']_{22} - [G']_{12} = -2A[\Delta G_{RSP}] \approx -2\Delta\alpha[\Delta G_{RSP}] \tag{20}$$

Here, we introduced the following notation.

$$[\Delta G_{RSP}] = \begin{pmatrix} 0 & 1 & f \\ 1 & 0 & 0 \\ f & 0 & 0 \end{pmatrix} \tag{21}$$

and

$$f = \frac{B}{A} = \frac{\Delta\gamma}{\Delta\alpha} \tag{22}$$

The eigenvalues and eigenvectors of $[\Delta G_{RSP}]$ can be written as $-\lambda_{RSP}, 0, \lambda_{RSP}$

$$\lambda_{RSP} = \sqrt{f^2 + 1} \tag{23}$$

Accordingly, the corresponding eigenvalues of $[G']_{22} - [G']_{12}$ are $-\lambda_{3\,RSP}, 0, \lambda_{3\,RSP}$ with

$$\lambda_{3\,RSP} = -2\Delta\alpha\lambda_{RSP} \tag{24}$$

It is straightforward to see that the orthogonal matrix of eigenvectors of $[\Delta G'_{RSP}]$ (as well as $[G']_{22} - [G']_{12}$) can be expressed as

$$[V_{RSP}] = \frac{1}{\sqrt{2}\lambda_{RSP}} \begin{pmatrix} -\lambda_{RSP} & 0 & \lambda_{RSP} \\ 1 & \sqrt{2}f & 1 \\ f & -\sqrt{2} & f \end{pmatrix} \tag{25}$$

Accordingly, two PDWs normal to the vectors $RSP_i^{(1,2)} \sim (V_{RSP\,i1} \mp V_{RSP\,i3})$ exist.

$$[RSP^{(1)}] = \begin{pmatrix} 1 \\ 0 \\ 0 \end{pmatrix}$$

$$[RSP^{(2)}] = \begin{pmatrix} 0 \\ 1 \\ f \end{pmatrix} \tag{26}$$

As in the case of PDWs connecting domain pairs of the type *R-sibling*, both *W*-and *S*-type DW are present. Notably, it is shown in equation (22) that the orientation of the DW depends on the ratio of the angles $\Delta\gamma$ and $\Delta\alpha$ rather than the lengths of the pseudocubic cell edges. The corresponding S-wall becomes parallel to the lattice plane with rational "Miller indices" for the special case such as $\Delta\gamma = \Delta\alpha, \Delta\alpha = 0$ or $\Delta\gamma = 0$. Table 4 lists these favorable cases.

**Table 4**  The same as Table 3 just for the case of S-walls separating the domain pairs of the "*R-semi-planar*" types.

| Lattice parameters | $f$ | $\lambda_{3\,RSP}$ | S-wall orientation |
| --- | --- | --- | --- |
| $\Delta\gamma = \Delta\alpha$ | 1 | $-2\sqrt{2}\Delta\alpha$ | (011) |
| $\Delta\alpha = 0$ | $\infty$ | $-2\Delta\gamma$ | (001) |
| $\Delta\gamma = 0$ | 0 | $-2\Delta\alpha$ | (010) |
| $\Delta\gamma = -\Delta\alpha$ | $\bar{1}$ | $-2\sqrt{2}\Delta\alpha$ | (01$\bar{1}$) |

### 3.4. PDWs connecting domain pairs of the type "*R-semi-crossed*".

We will demonstrate the derivation of the PDWs connecting the representative domain pair $M_{12}\,M_{23}$ and obtain the similar results for all the other pairs of this type analogously. Using the last column of Table 1 and equations (6)

$$[G']_{23} - [G']_{12} = -(A+B)[\Delta G_{RSC}] \approx -(\Delta\alpha + \Delta\gamma)[\Delta G_{RSC}] \tag{27}$$

Here we introduced the following notation.

$$[\Delta G_{RSC}] = \begin{pmatrix} 0 & 1 & 1 \\ 1 & p & 0 \\ 1 & 0 & \bar{p} \end{pmatrix} \tag{28}$$

and

$$p = \frac{C}{A+B} \approx \frac{2}{\Delta\alpha + \Delta\gamma}\left(\frac{c}{a} - 1\right) \tag{29}$$

The eigenvectors and eigenvalues of the $[\Delta G_{RSC}]$ can be found as $-\lambda_{RSC}, 0, \lambda_{RSC}$

$$\lambda_{RSC} = \sqrt{p^2 + 2} \tag{30}$$

Accordingly, the corresponding eigenvalues of $[G']_{23} - [G']_{12}$ are $-\lambda_{3\,RSC}, 0, \lambda_{3\,RSC}$ with

$$\lambda_{3\,RSC} = -(\Delta\alpha + \Delta\gamma)\lambda_{RSC} \tag{31}$$

It is straightforward to see that the orthogonal matrix of eigenvectors of $[\Delta G_{RSC}]$ (as well as $[G']_{23} - [G']_{12}$) can be expressed as

$$[V_{RSC}] = \frac{1}{2\lambda_{RSC}} \begin{pmatrix} \bar{2} & 2p & 2 \\ \lambda_{RSC} - p & \bar{2} & \lambda_{RSC} + p \\ \lambda_{RSC} + p & 2 & \lambda_{RSC} - p \end{pmatrix} \tag{32}$$

Accordingly, two PDWs normal to the vectors $RSC_i^{(1,2)} \sim (V_{RSC\,i1} \mp V_{RSC\,i3})$ exist.

$$[RSC^{(1)}] = \begin{pmatrix} 2 \\ p \\ \bar{p} \end{pmatrix} \tag{33}$$

$$[RSC^{(2)}] = \begin{pmatrix} 0 \\ 1 \\ 1 \end{pmatrix}$$

As for the cases of DWs connecting domain pairs of the type "*R-sibling*" and "*R-semi-planar*", $W$-and $S$-type DWs are present here. In addition, some favorable cases (Table 5) of the lattice parameters turn the S-type of PDW into PDW with rational Miller indices.

**Table 5**   The same as Table 3 just for the case of S-walls separating the domain pairs of the "*R-semi-crossed*" type.

| Lattice parameters | $p$ | $\lambda_{3\,RSC}$ | S-wall orientation |
|---|---|---|---|
| $\left(\dfrac{c}{a} - 1\right) = \Delta\alpha + \Delta\gamma$ | 2 | $-(\Delta\alpha + \Delta\gamma)\sqrt{6}$ | $(11\bar{1})$ |
| $a = c$ | 0 | $-(\Delta\alpha + \Delta\gamma)\sqrt{2}$ | $(100)$ |
| $\Delta\gamma = -\Delta\alpha$ | $\infty$ | $2\left(1 - \dfrac{c}{a}\right)$ | $(01\bar{1})$ |

### 3.5. PDWs connecting domains pairs of the type "*R-crossed*".

We will show that the corresponding domain pairs of this type do not generally have any PDWs. Indeed, we can attempt to find such for the case of the representative pair of domains $M_{32}\,M_{23}$. According to the last column of Tabel 1 and equations (*6*) we get:

$$[G']_{32} - [G']_{23} = \begin{pmatrix} 0 & B - A & B + A \\ B - A & C & -2A \\ B + A & -2A & \bar{C} \end{pmatrix} = [\Delta G_{RC}] \tag{34}$$

The determinant of $[\Delta G_{RC}]$ can be calculated as

$$|\Delta G_{RC}| = 4A(A^2 - B^2 - BC) \tag{35}$$

Accordingly, this pair of domains may connect along the PDW if one of the following conditions is fulfilled:

$$A = 0 \ \text{ or } \ \Delta\alpha = 0 \tag{36}$$

or

$$A^2 - B^2 - BC = 0 \tag{37}$$

These conditions are generally not fulfilled and therefore we can consider domain pairs of the type "crossed" not compatible. The special conditions under which domain pairs may connect can be the subject of future work.

#### 4. The change of the polarization direction across the domain walls

The SPD changes across any DW. This paragraph demonstrates the calculation of the change of the SPD projection on the DW normal. Such a change is numerically equal to the surface density of electric charge at the wall (Jackson, 2007). We will consider that each ferroelastic domain $mn$ may host spontaneous polarization $+\boldsymbol{P}_{mn}$ or $-\boldsymbol{P}_{mn}$ (the coordinates of the vectors $\boldsymbol{P}_{mn}$ are defined in Table 1). Accordingly, the specific pair of domains $m_1 n_1$ and $m_2 n_2$ may meet along a DW that switches SPD according to configuration $\boldsymbol{P}_{m_1 n_1} \rightarrow \boldsymbol{P}_{m_2 n_2}$ (+) or $\boldsymbol{P}_{m_1 n_1} \rightarrow -\boldsymbol{P}_{m_2 n_2}$ (−). We will see which of these configurations ensures zero (or minimal) charge at the corresponding DW using equations (14), (19), (26), (33) for the normal to the DW of each type. Table 6 summarizes the results. It shows that uncharged DWs occur in the following cases:

- *W*- and *S*-type "*R-sibling*" PDWs change SPD by nearly 180 or 0°, respectively.

- (100)- and (011) - "*R-planar*" PDWs change SPD by 109 and 71°, respectively.

- *W*- and *S*-type "*R-semi-planar*" PDWs change SPD by 109 and 71°, respectively.

- *W*-and *S*-type "*R-semi-crossed*" PDWs change SPD by 71 and 109°, respectively.

These results hold significant implications, particularly in the context of describing the DW motion under external electric fields and assessing the role of the specifically connected domain pair to in the extrinsic contribution to the electromechanical coupling (Pramanick *et al.*, 2011; Jones *et al.*, 2006; Tutuncu *et al.*, 2016; Gorfman *et al.*, 2020). Indeed, this contribution hinges on the orientation of the SPD with respect to electric field: domains with positive / negative projection of the SPD to the applied electric field would expand / contract, respectively. Consequently, comprehending the SPD's orientation and its change across the DW is pivotal.

**Table 6**     The change of the SPD across each of the DWs described above. The first column contains the type of domain pair. The second column contains the PDW "Miller" indices. The third column contains the domain numbers $m_1 n_1 \mid m_2 n_2$ meeting along the wall. The fourth column contains the sign involved in the connection: the sign + means e.g. $\boldsymbol{P}_{m_1 n_1} \rightarrow \boldsymbol{P}_{m_2 n_2}$, the sign − stands for e.g., $\boldsymbol{P}_{m_1 n_1} \rightarrow -\boldsymbol{P}_{m_2 n_2}$. The fifth column contains the projection of the $\boldsymbol{P}_{m_1 n_1}$ to the DW normal. The last column shows the angles between the corresponding SPD as defined in Table 2 and at $\rho = 0$.

| Type | Orientation | Domain pair | Sign | Projection | $\xi_0$ (°) |
|------|-------------|-------------|------|------------|-------------|
| RSB | $(0\bar{1}1)$ | 12  \|  13 | − | $1 - x$ | 180 |
| RSB | $(2rr)$ | 12  \|  13 | + | $2r + 2rx$ | 0 |
| RP | (100) | 11  \|  21 | − | $x$ | 109 |
| RP | (011) | 11  \|  21 | + | 2 | 71 |

| RSP | (100) | 12 \| 22 | − | 1 | 109 |
| RSP | ($01f$) | 12 \| 22 | + | $f$ | 71 |
| RSC | ($2p\bar{p}$) | 12 \| 23 | − | $2 + px - p$ | 109 |
| RSC | (011) | 12 \| 23 | + | $1 + x$ | 71 |

## 5. Derivation of the transformation matrices and the separation between Bragg peaks

### 5.1. General expressions

After calculating the indices of PDWs, connecting the specific pair of domains, it is also possible to calculate the orientation relationship between domains and separation of Bragg peaks diffracted from them. Full details of these calculations are presented by Gorfman *et al.* (2022) and briefly summarized here. The matrix of transformation $[\Delta S]$ between the basis vectors of the domains $m$ and $n$ is defined as $(\boldsymbol{a}_{1n} \quad \boldsymbol{a}_{2n} \quad \boldsymbol{a}_{3n}) = (\boldsymbol{a}_{1m} \quad \boldsymbol{a}_{2m} \quad \boldsymbol{a}_{3m})([I] + [\Delta S])$. This matrix can be calculated according to:

$$[\Delta S] = [V][Z]\begin{pmatrix} 0 & 0 & y_1 \\ 0 & 0 & y_2 \\ 0 & 0 & 0 \end{pmatrix}[Z]^{-1}[V]^T \tag{38}$$

Here

$$[Z] = \begin{pmatrix} 1 & 0 & 0 \\ 0 & 1 & 0 \\ \pm 1 & 0 & 1 \end{pmatrix} \tag{39}$$

The sign $\pm$ before $Z_{31}$ is used for the cases when PDW normal is $V_{i1} \mp V_{i3}$ respectively. The coefficients $y_1$ and $y_2$ can be calculated according to:

$$\begin{pmatrix} G_{m,11}^{(W)} & G_{m,12}^{(W)} \\ G_{m,21}^{(W)} & G_{m,22}^{(W)} \end{pmatrix}\begin{pmatrix} y_1 \\ y_2 \end{pmatrix} = \begin{pmatrix} G_{n,13}^{(W)} - G_{m,13}^{(W)} \\ G_{n,23}^{(W)} - G_{m,23}^{(W)} \end{pmatrix} \tag{40}$$

With $\left[G_{m,n}^{(W)}\right]$ being defined as

$$\left[G_{m,n}^{(W)}\right] = [Z]^T[Z] + [Z]^T[V]^T\left[G_{m,n}'\right][V][Z] \tag{41}$$

Similarly, the matrix of transformation $[\Delta S^*]$ between the reciprocal basis vectors of the domains $m$ and $n$ is defined as $(\boldsymbol{a}_{1n}^* \quad \boldsymbol{a}_{2n}^* \quad \boldsymbol{a}_{3n}^*) = (\boldsymbol{a}_{1m}^* \quad \boldsymbol{a}_{2m}^* \quad \boldsymbol{a}_{3m}^*)([I] + [\Delta S^*])$ and can be calculated according to:

$$[\Delta S^*]^T = [V][Z]\begin{pmatrix} 0 & 0 & \bar{y}_1 \\ 0 & 0 & \bar{y}_2 \\ 0 & 0 & 0 \end{pmatrix}[Z]^{-1}[V]^T \tag{42}$$

We can use (42) to calculate the separation between the Bragg peaks $H, K, L$ so that:

$$[\Delta \boldsymbol{B}] = \begin{pmatrix} \Delta H \\ \Delta K \\ \Delta L \end{pmatrix} = [\Delta S^*] \begin{pmatrix} H \\ K \\ L \end{pmatrix} \tag{43}$$

## 5.2. Simplifications

Equations (38) and (42) can be used to obtain the elements of $[\Delta S]$ and $[\Delta S^*]$ numerically. However, we will show that reasonable approximation leads to more visually appealing analytical expressions. Let us notice that the right-hand side of (40) can be derived from the elements 13 and 23 of the matrix $[\Delta G^{(W)}] = [Z]^T [V]^T [\Delta G'][V][Z]$. Considering that the columns of the matrix $[V]$ are the eigenvectors of $[\Delta G']$ with the eigenvectors of $-\lambda_3$, $0$ and $\lambda_3$ respectively we can write.

$$[\Delta G^{(W)}] = \lambda_3 [Z]^T \begin{pmatrix} \bar{1} & 0 & 0 \\ 0 & 0 & 0 \\ 0 & 0 & 1 \end{pmatrix} [Z] = \lambda_3 \begin{pmatrix} 0 & 0 & \pm 1 \\ 0 & 0 & 0 \\ \pm 1 & 0 & 1 \end{pmatrix} \tag{44}$$

Here, the same sign as in the equation (39) is implemented instead of $\pm$. Using (44) we can rewrite (40) as:

$$\begin{pmatrix} G_{m,11}^{(W)} & G_{m,12}^{(W)} \\ G_{m,21}^{(W)} & G_{m,22}^{(W)} \end{pmatrix} \begin{pmatrix} y_1 \\ y_2 \end{pmatrix} = \pm \lambda_3 \begin{pmatrix} 1 \\ 0 \end{pmatrix} \tag{45}$$

We will now consider that the second term in the right side of equation (41) $[Z]^T [V]^T [G'_{m,n}][V][Z]$ is proportional to the parameters of monoclinic distortion $A, B, C$ (see equations (6)) and therefore it is much smaller than the first term $[Z]^T [Z]$. Accordingly, we can rewrite (41) in the form:

$$[G^{(W)}]_m = [Z]^T [Z] + O(A, B, C) \approx \begin{pmatrix} 2 & 0 & \pm 1 \\ 0 & 1 & 0 \\ \pm 1 & 0 & 1 \end{pmatrix} \tag{46}$$

Considering (46) we can re-write (45) as:

$$\begin{aligned} y_1 &= \pm \frac{\lambda_3}{2} \\ y_2 &= 0 \end{aligned} \tag{47}$$

Substituting (47) into (38) and (42) we get

$$[\Delta S] = \pm \frac{\lambda_3}{2} [V] \begin{pmatrix} 1 \\ 0 \\ \pm 1 \end{pmatrix} (\mp 1 \quad 0 \quad 1)[V]^T \tag{48}$$

and

$$[\Delta S^*] = \pm \frac{\lambda_3}{2} [V] \begin{pmatrix} 1 \\ 0 \\ \mp 1 \end{pmatrix} (\pm 1 \quad 0 \quad 1)[V]^T \tag{49}$$

Using the notations $[\Delta S_+]$, $[\Delta S_-]$ for the case of signs + and − in front of $\lambda_3$ respectively, we can see that (48), (49) lead to:

$$\begin{aligned}
[\Delta S_-] &= [\Delta S_+]^T \\
[\Delta S_+^*] &= -[\Delta S_-] \\
[\Delta S_-^*] &= -[\Delta S_+]
\end{aligned} \tag{50}$$

### 5.3. The case of domain pairs of the type "*R-sibling*"

We will now apply (48) and (49) for PDWs connecting domain pairs of the type *R-sibling*. The corresponding transformation matrices $[\Delta S_+]$, $[\Delta S_-]$ are marked explicitly as $\left[\Delta S_{(0\bar{1}1)}^{RSB}\right]$ and $\left[\Delta S_{(2rr)}^{RSB}\right]$. According to (12) we obtain $\lambda_3 = (\alpha - \gamma)\sqrt{2 + r^2}$. Substituting (13) into (48) and using (50) we get:

$$\left[\Delta S_{(0\bar{1}1)}^{RSB}\right] = \left[\Delta S_{(2rr)}^{RSB}\right]^T = \frac{\alpha - \gamma}{2}\begin{pmatrix} 0 & 2 & \bar{2} \\ 0 & r & \bar{r} \\ 0 & r & \bar{r} \end{pmatrix} \tag{51}$$

Using (43) and (50) we can obtain the separation between the Bragg peaks diffracted from the corresponding pair of domains as

$$\left[\Delta \boldsymbol{B}_{(0\bar{1}1)}^{RSB}\right] = \frac{\alpha - \gamma}{2}(2H + rK + rL)\begin{pmatrix} 0 \\ \bar{1} \\ 1 \end{pmatrix} \tag{52}$$

and

$$\left[\Delta \boldsymbol{B}_{(2rr)}^{RSB}\right] = \frac{\alpha - \gamma}{2}(L - K)\begin{pmatrix} 2 \\ r \\ r \end{pmatrix} \tag{53}$$

As mentioned by Gorfman *et al.* (2022) the three-dimensional separation between the Bragg peaks diffracted from pair of connected domains is parallel to the DW normal.

### 5.4. The case of domain pairs of the type "*R-planar*"

Similarly, for the case of domain pairs of the type "R-planar", the corresponding transformation matrices $[\Delta S_+]$, $[\Delta S_-]$ are marked explicitly as $\left[\Delta S_{(100)}^{RP}\right]$ and $\left[\Delta S_{(011)}^{RP}\right]$. According to (17) $\lambda_3 = -2\sqrt{2}\Delta\alpha$. Substituting (18) into (48) and using (50) we obtain:

$$\left[\Delta S_{(100)}^{RP}\right] = \left[\Delta S_{(011)}^{RP}\right]^T = -2\Delta\alpha\begin{pmatrix} 0 & 0 & 0 \\ 1 & 0 & 0 \\ 1 & 0 & 0 \end{pmatrix} \tag{54}$$

Equivalently we will obtain the following expression for the separation of Bragg peaks

$$\left[\Delta \boldsymbol{B}_{(100)}^{RP}\right] = 2\Delta\alpha(K + L)\begin{pmatrix} 1 \\ 0 \\ 0 \end{pmatrix} \tag{55}$$

and

$$\left[\Delta\boldsymbol{B}^{RP}_{(011)}\right] = 2\Delta\alpha H\begin{pmatrix}0\\1\\1\end{pmatrix} \tag{56}$$

## 5.5. The case of domain pairs of the type "*R-semi-planar*"

The corresponding transformation matrices $[\Delta S_+], [\Delta S_-]$ are then marked explicitly as $\left[\Delta S^{RSP}_{(100)}\right]$ and $\left[\Delta S^{RSP}_{(01f)}\right]$. According to (24) $\lambda_3 = -2\Delta\alpha\sqrt{f^2+1}$. Substituting (25) into (48) and using (50):

$$\left[\Delta S^{RSP}_{(100)}\right] = \left[\Delta S^{RSP}_{(01f)}\right]^T = -2\Delta\alpha\begin{pmatrix}0&0&0\\1&0&0\\f&0&0\end{pmatrix} \tag{57}$$

The separation of the Bragg peaks diffracted from the correspondingly connected domain pairs are

$$\left[\Delta\boldsymbol{B}^{RSP}_{(100)}\right] = 2\Delta\alpha(K+fL)\begin{pmatrix}1\\0\\0\end{pmatrix} \tag{58}$$

and

$$\left[\Delta\boldsymbol{B}^{RSP}_{(01f)}\right] = 2\Delta\alpha H\begin{pmatrix}0\\1\\f\end{pmatrix} \tag{59}$$

## 5.6. The case of domain pairs of the type "*R-semi-crossed*"

The corresponding transformation matrices $[\Delta S_+], [\Delta S_-]$ are marked explicitly as $\left[\Delta S^{RSC}_{(2p\bar{p})}\right]$ and $\left[\Delta S^{RSC}_{(011)}\right]$. According to (31) $\lambda_3 = -(\Delta\alpha + \Delta\gamma)\sqrt{2+p^2}$. Substituting (32) into (48) and using (50):

$$\left[\Delta S^{RSC}_{(2p\bar{p})}\right] = \left[\Delta S^{RSC}_{(011)}\right]^T = -\frac{\Delta\gamma+\Delta\alpha}{2}\begin{pmatrix}0&0&0\\2&p&\bar{p}\\2&p&\bar{p}\end{pmatrix} \tag{60}$$

The separation of the Bragg peaks diffracted from the domains, meeting along the DW normal to $[RSC^{(1)}]$

$$\left[\Delta\boldsymbol{B}^{RSC}_{(2p\bar{p})}\right] = (\Delta\gamma+\Delta\alpha)(K+L)\begin{pmatrix}2\\p\\\bar{p}\end{pmatrix} \tag{61}$$

And for the case of $[RSC^{(2)}]$

$$\left[\Delta\boldsymbol{B}^{RSC}_{(011)}\right] = (\Delta\gamma+\Delta\alpha)(2H+pK-pL)\begin{pmatrix}0\\1\\1\end{pmatrix} \tag{62}$$

## 5.7. Summarizing tables

The previous paragraphs demonstrated how to derive key quantities such as Miller indices, orientation relationship between the lattice basis vectors, and the separation of Bragg peaks for representative domain pairs only. Similar equations can be derived for all the other domain pairs. The tables and figures below list the corresponding quantities for all 84 existing PDWs. Their full list includes:

- 24 PDWs connecting domain pairs of the type "*R-sibling*", including 12 W- and 12 S-walls.

- 12 PDWs connecting domain pairs of the type "*R-planar*". All of them are W-walls.

- 24 PDWs connecting domain pairs of the type "*R-semi-planar*", including 12 W-walls and 12 S walls.

- 24 PDWs connecting domain pairs of the type "*R-semi-crossed*", including 12 W-walls and 12 S-walls.

The list of 84 PDWs contains 36 *S*-walls and 48 *W*-walls as listed in Tables 7, 8, 9, 10. Each row of these tables contains domain pair number, the Miller indices of the PDW, the matrix of transformation $[\Delta S]$ between the corresponding basis vectors and the separation of Bragg peaks $H, K, L$ diffracted from this pair of domains.

**Table 7**  Summary of 24 PDW connecting domain pairs of the "*R-sibling*" type.  The first column contains the PDW number, while the second and the third columns contain the domain identifiers based on Figure 1 and Table 1. The fourth column displays the "Miller" indices of the PDW. The fifth column contains the transformation matrix between the basis vectors of the domain $m_1 n_1$ to the basis vectors of the domain $m_2 n_2$. The last column contains the separation between the Bragg peak with the indices $H, K, L$ diffracted from these domains.

| $N$ | $M_{m_1 n_1}$ | $M_{m_2 n_2}$ | $(hkl)$ | $[\Delta S]\ (\frac{\alpha-\gamma}{2})$ | $[\Delta \boldsymbol{B}]\ (\frac{\alpha-\gamma}{2})$ |
|---|---|---|---|---|---|
| 1 | $M_{11}$ | $M_{12}$ | $(\bar{1}\,1\,0)$ | $\begin{pmatrix} r & \bar{r} & 0 \\ r & \bar{r} & 0 \\ 2 & \bar{2} & 0 \end{pmatrix}$ | $(rH + rK + 2L)\begin{pmatrix} \bar{1} \\ 1 \\ 0 \end{pmatrix}$ |
| 2 | $M_{11}$ | $M_{12}$ | $(r\,r\,2)$ | $\begin{pmatrix} r & r & 2 \\ \bar{r} & \bar{r} & \bar{2} \\ 0 & 0 & 0 \end{pmatrix}$ | $(K + \bar{H})\begin{pmatrix} r \\ r \\ 2 \end{pmatrix}$ |
| 3 | $M_{11}$ | $M_{13}$ | $(\bar{1}\,0\,1)$ | $\begin{pmatrix} r & 0 & \bar{r} \\ 2 & 0 & \bar{2} \\ r & 0 & \bar{r} \end{pmatrix}$ | $(rH + 2K + rL)\begin{pmatrix} \bar{1} \\ 0 \\ 1 \end{pmatrix}$ |
| 4 | $M_{11}$ | $M_{13}$ | $(r\,2\,r)$ | $\begin{pmatrix} r & 2 & r \\ 0 & 0 & 0 \\ \bar{r} & \bar{2} & \bar{r} \end{pmatrix}$ | $(L + \bar{H})\begin{pmatrix} r \\ 2 \\ r \end{pmatrix}$ |
| 5 | $M_{12}$ | $M_{13}$ | $(0\,\bar{1}\,1)$ | $\begin{pmatrix} 0 & 2 & \bar{2} \\ 0 & r & \bar{r} \\ 0 & r & \bar{r} \end{pmatrix}$ | $(2H + rK + rL)\begin{pmatrix} 0 \\ \bar{1} \\ 1 \end{pmatrix}$ |
| 6 | $M_{12}$ | $M_{13}$ | $(2\,r\,r)$ | $\begin{pmatrix}0&0&0\\2&r&r\\\bar{2}&\bar{r}&\bar{r}\end{pmatrix}$ | $(L+\bar{K})\begin{pmatrix}2\\r\\r\end{pmatrix}$ |
| 7 | $M_{21}$ | $M_{22}$ | $(1\,1\,0)$ | $\begin{pmatrix}r&r&0\\\bar{r}&\bar{r}&0\\\bar{2}&\bar{2}&0\end{pmatrix}$ | $(r\bar{H}+rK+2L)\begin{pmatrix}1\\1\\0\end{pmatrix}$ |
| 8 | $M_{21}$ | $M_{22}$ | $(\bar{r}\,r\,2)$ | $\begin{pmatrix}r&\bar{r}&\bar{2}\\r&\bar{r}&\bar{2}\\0&0&0\end{pmatrix}$ | $(H+K)\begin{pmatrix}\bar{r}\\r\\2\end{pmatrix}$ |
| 9 | $M_{21}$ | $M_{23}$ | $(1\,0\,1)$ | $\begin{pmatrix}r&0&r\\\bar{2}&0&\bar{2}\\\bar{r}&0&\bar{r}\end{pmatrix}$ | $(r\bar{H}+2K+rL)\begin{pmatrix}1\\0\\1\end{pmatrix}$ |
| 10 | $M_{21}$ | $M_{23}$ | $(\bar{r}\,2\,r)$ | $\begin{pmatrix}r&\bar{2}&\bar{r}\\0&0&0\\r&\bar{2}&\bar{r}\end{pmatrix}$ | $(H+L)\begin{pmatrix}\bar{r}\\2\\r\end{pmatrix}$ |
| 11 | $M_{22}$ | $M_{23}$ | $(0\,\bar{1}\,1)$ | $\begin{pmatrix}0&\bar{2}&2\\0&r&\bar{r}\\0&r&\bar{r}\end{pmatrix}$ | $(2\bar{H}+rK+rL)\begin{pmatrix}0\\\bar{1}\\1\end{pmatrix}$ |
| 12 | $M_{22}$ | $M_{23}$ | $(\bar{2}\,r\,r)$ | $\begin{pmatrix}0&0&0\\\bar{2}&r&r\\2&\bar{r}&\bar{r}\end{pmatrix}$ | $(L+\bar{K})\begin{pmatrix}\bar{2}\\r\\r\end{pmatrix}$ |
| 13 | $M_{31}$ | $M_{32}$ | $(1\,1\,0)$ | $\begin{pmatrix}r&r&0\\\bar{r}&\bar{r}&0\\2&2&0\end{pmatrix}$ | $(r\bar{H}+rK+2\bar{L})\begin{pmatrix}1\\1\\0\end{pmatrix}$ |
| 14 | $M_{31}$ | $M_{32}$ | $(r\,\bar{r}\,2)$ | $\begin{pmatrix}r&\bar{r}&2\\r&\bar{r}&2\\0&0&0\end{pmatrix}$ | $(\bar{H}+\bar{L})\begin{pmatrix}r\\\bar{r}\\2\end{pmatrix}$ |
| 15 | $M_{31}$ | $M_{33}$ | $(1\,0\,\bar{1})$ | $\begin{pmatrix}r&0&\bar{r}\\\bar{2}&0&2\\r&0&\bar{r}\end{pmatrix}$ | $(r\bar{H}+2K+r\bar{L})\begin{pmatrix}1\\0\\\bar{1}\end{pmatrix}$ |
| 16 | $M_{31}$ | $M_{33}$ | $(r\,\bar{2}\,r)$ | $\begin{pmatrix}r&\bar{2}&r\\0&0&0\\\bar{r}&2&\bar{r}\end{pmatrix}$ | $(\bar{H}+L)\begin{pmatrix}r\\\bar{2}\\r\end{pmatrix}$ |
| 17 | $M_{32}$ | $M_{33}$ | $(0\,1\,1)$ | $\begin{pmatrix}0&\bar{2}&\bar{2}\\0&r&r\\0&\bar{r}&\bar{r}\end{pmatrix}$ | $(2H+r\bar{K}+rL)\begin{pmatrix}0\\1\\1\end{pmatrix}$ |
| 18 | $M_{32}$ | $M_{33}$ | $(2\,\bar{r}\,r)$ | $\begin{pmatrix}0&0&0\\\bar{2}&r&\bar{r}\\\bar{2}&r&\bar{r}\end{pmatrix}$ | $(K+L)\begin{pmatrix}2\\\bar{r}\\r\end{pmatrix}$ |
| 19 | $M_{41}$ | $M_{42}$ | $(\bar{1}\,1\,0)$ | $\begin{pmatrix}r&\bar{r}&0\\r&\bar{r}&0\\\bar{2}&2&0\end{pmatrix}$ | $(rH+rK+2\bar{L})\begin{pmatrix}\bar{1}\\1\\0\end{pmatrix}$ |

| N | $M_{m_1 n_1}$ | $M_{m_2 n_2}$ | $(hkl)$ | $[\Delta S]$ $(2\Delta\alpha)$ | $[\Delta \boldsymbol{B}]$ $(2\Delta\alpha)$ |
|---|---|---|---|---|---|
| 20 | $M_{41}$ | $M_{42}$ | $(\bar r\,\bar r\,2)$ | $\begin{pmatrix} r & r & \bar2 \\ \bar r & \bar r & 2 \\ 0 & 0 & 0 \end{pmatrix}$ | $(H+\bar K)\begin{pmatrix}\bar r\\\bar r\\2\end{pmatrix}$ |
| 21 | $M_{41}$ | $M_{43}$ | $(1\,0\,1)$ | $\begin{pmatrix} r & 0 & r \\ 2 & 0 & 2 \\ \bar r & 0 & \bar r \end{pmatrix}$ | $(r\bar H+2\bar K+rL)\begin{pmatrix}1\\0\\1\end{pmatrix}$ |
| 22 | $M_{41}$ | $M_{43}$ | $(\bar r\,\bar 2\,r)$ | $\begin{pmatrix} r & 2 & \bar r \\ 0 & 0 & 0 \\ r & 2 & \bar r \end{pmatrix}$ | $(H+L)\begin{pmatrix}\bar r\\\bar2\\r\end{pmatrix}$ |
| 23 | $M_{42}$ | $M_{43}$ | $(0\,1\,1)$ | $\begin{pmatrix} 0 & 2 & 2 \\ 0 & r & r \\ 0 & \bar r & \bar r \end{pmatrix}$ | $(2\bar H+r\bar K+rL)\begin{pmatrix}0\\1\\1\end{pmatrix}$ |
| 24 | $M_{42}$ | $M_{43}$ | $(\bar 2\,\bar r\,r)$ | $\begin{pmatrix} 0 & 0 & 0 \\ 2 & r & \bar r \\ 2 & r & \bar r \end{pmatrix}$ | $(K+L)\begin{pmatrix}\bar2\\\bar r\\r\end{pmatrix}$ |

**Table 8**  The same as Table 7 but for the case of PDWs, connecting domain pairs of the type "*R-planar*".

| N | $M_{m_1 n_1}$ | $M_{m_2 n_2}$ | $(hkl)$ | $[\Delta S]$ $(2\Delta\alpha)$ | $[\Delta \boldsymbol{B}]$ $(2\Delta\alpha)$ |
|---|---|---|---|---|---|
| 25 | $M_{11}$ | $M_{21}$ | $(1\,0\,0)$ | $\begin{pmatrix} 0 & 0 & 0 \\ \bar1 & 0 & 0 \\ \bar1 & 0 & 0 \end{pmatrix}$ | $(K+L)\begin{pmatrix}1\\0\\0\end{pmatrix}$ |
| 26 | $M_{11}$ | $M_{21}$ | $(0\,1\,1)$ | $\begin{pmatrix} 0 & \bar1 & \bar1 \\ 0 & 0 & 0 \\ 0 & 0 & 0 \end{pmatrix}$ | $H\begin{pmatrix}0\\1\\1\end{pmatrix}$ |
| 27 | $M_{12}$ | $M_{32}$ | $(0\,1\,0)$ | $\begin{pmatrix} 0 & \bar1 & 0 \\ 0 & 0 & 0 \\ 0 & \bar1 & 0 \end{pmatrix}$ | $(H+L)\begin{pmatrix}0\\1\\0\end{pmatrix}$ |
| 28 | $M_{12}$ | $M_{32}$ | $(1\,0\,1)$ | $\begin{pmatrix} 0 & 0 & 0 \\ \bar1 & 0 & \bar1 \\ 0 & 0 & 0 \end{pmatrix}$ | $K\begin{pmatrix}1\\0\\1\end{pmatrix}$ |
| 29 | $M_{13}$ | $M_{43}$ | $(0\,0\,1)$ | $\begin{pmatrix} 0 & 0 & \bar1 \\ 0 & 0 & \bar1 \\ 0 & 0 & 0 \end{pmatrix}$ | $(H+K)\begin{pmatrix}0\\0\\1\end{pmatrix}$ |
| 30 | $M_{13}$ | $M_{43}$ | $(1\,1\,0)$ | $\begin{pmatrix} 0 & 0 & 0 \\ 0 & 0 & 0 \\ \bar1 & \bar1 & 0 \end{pmatrix}$ | $L\begin{pmatrix}1\\1\\0\end{pmatrix}$ |
| 31 | $M_{23}$ | $M_{33}$ | $(0\,0\,1)$ | $\begin{pmatrix} 0 & 0 & 1 \\ 0 & 0 & \bar1 \\ 0 & 0 & 0 \end{pmatrix}$ | $(\bar H+K)\begin{pmatrix}0\\0\\1\end{pmatrix}$ |

| $N$ | $M_{m_1 n_1}$ | $M_{m_2 n_2}$ | $(hkl)$ | $[\Delta S]$ $(2\Delta\alpha)$ | $[\Delta B]$ $(2\Delta\alpha)$ |
|---|---|---|---|---|---|
| 32 | $M_{23}$ | $M_{33}$ | $(\bar{1}\,1\,0)$ | $\begin{pmatrix} 0 & 0 & 0 \\ 0 & 0 & 0 \\ 1 & \bar{1} & 0 \end{pmatrix}$ | $L\begin{pmatrix} \bar{1} \\ 1 \\ 0 \end{pmatrix}$ |
| 33 | $M_{22}$ | $M_{42}$ | $(0\,1\,0)$ | $\begin{pmatrix} 0 & 1 & 0 \\ 0 & 0 & 0 \\ 0 & \bar{1} & 0 \end{pmatrix}$ | $(\bar{H}+L)\begin{pmatrix} 0 \\ 1 \\ 0 \end{pmatrix}$ |
| 34 | $M_{22}$ | $M_{42}$ | $(\bar{1}\,0\,1)$ | $\begin{pmatrix} 0 & 0 & 0 \\ 1 & 0 & \bar{1} \\ 0 & 0 & 0 \end{pmatrix}$ | $K\begin{pmatrix} \bar{1} \\ 0 \\ 1 \end{pmatrix}$ |
| 35 | $M_{31}$ | $M_{41}$ | $(1\,0\,0)$ | $\begin{pmatrix} 0 & 0 & 0 \\ 1 & 0 & 0 \\ \bar{1} & 0 & 0 \end{pmatrix}$ | $(\bar{K}+L)\begin{pmatrix} 1 \\ 0 \\ 0 \end{pmatrix}$ |
| 36 | $M_{31}$ | $M_{41}$ | $(0\,\bar{1}\,1)$ | $\begin{pmatrix} 0 & 1 & \bar{1} \\ 0 & 0 & 0 \\ 0 & 0 & 0 \end{pmatrix}$ | $H\begin{pmatrix} 0 \\ \bar{1} \\ 1 \end{pmatrix}$ |

**Table 9** The same as Table 7 but for the case of PDWs, connecting domain pairs of the type "*R-semi-planar*".

| $N$ | $M_{m_1 n_1}$ | $M_{m_2 n_2}$ | $(hkl)$ | $[\Delta S]$ $(2\Delta\alpha)$ | $[\Delta B]$ $(2\Delta\alpha)$ |
|---|---|---|---|---|---|
| 37 | $M_{12}$ | $M_{22}$ | $(1\,0\,0)$ | $\begin{pmatrix} 0 & 0 & 0 \\ \bar{1} & 0 & 0 \\ \bar{f} & 0 & 0 \end{pmatrix}$ | $(K+fL)\begin{pmatrix} 1 \\ 0 \\ 0 \end{pmatrix}$ |
| 38 | $M_{12}$ | $M_{22}$ | $(0\,1\,f)$ | $\begin{pmatrix} 0 & \bar{1} & \bar{f} \\ 0 & 0 & 0 \\ 0 & 0 & 0 \end{pmatrix}$ | $H\begin{pmatrix} 0 \\ 1 \\ f \end{pmatrix}$ |
| 39 | $M_{13}$ | $M_{23}$ | $(1\,0\,0)$ | $\begin{pmatrix} 0 & 0 & 0 \\ \bar{f} & 0 & 0 \\ \bar{1} & 0 & 0 \end{pmatrix}$ | $(fK+L)\begin{pmatrix} 1 \\ 0 \\ 0 \end{pmatrix}$ |
| 40 | $M_{13}$ | $M_{23}$ | $(0\,f\,1)$ | $\begin{pmatrix} 0 & \bar{f} & \bar{1} \\ 0 & 0 & 0 \\ 0 & 0 & 0 \end{pmatrix}$ | $H\begin{pmatrix} 0 \\ f \\ 1 \end{pmatrix}$ |
| 41 | $M_{11}$ | $M_{31}$ | $(0\,1\,0)$ | $\begin{pmatrix} 0 & \bar{1} & 0 \\ 0 & 0 & 0 \\ 0 & \bar{f} & 0 \end{pmatrix}$ | $(H+fL)\begin{pmatrix} 0 \\ 1 \\ 0 \end{pmatrix}$ |
| 42 | $M_{11}$ | $M_{31}$ | $(1\,0\,f)$ | $\begin{pmatrix} 0 & 0 & 0 \\ \bar{1} & 0 & \bar{f} \\ 0 & 0 & 0 \end{pmatrix}$ | $K\begin{pmatrix} 1 \\ 0 \\ f \end{pmatrix}$ |
| 43 | $M_{13}$ | $M_{33}$ | $(0\,1\,0)$ | $\begin{pmatrix} 0 & \bar{f} & 0 \\ 0 & 0 & 0 \\ 0 & \bar{1} & 0 \end{pmatrix}$ | $(fH+L)\begin{pmatrix} 0 \\ 1 \\ 0 \end{pmatrix}$ |
| 44 | $M_{13}$ | $M_{33}$ | $(f\,0\,1)$ | $\begin{pmatrix} 0 & 0 & 0 \\ \bar{f} & 0 & \bar{1} \\ 0 & 0 & 0 \end{pmatrix}$ | $K\begin{pmatrix} f \\ 0 \\ 1 \end{pmatrix}$ |
| 45 | $M_{11}$ | $M_{41}$ | $(0\,0\,1)$ | $\begin{pmatrix} 0 & 0 & \bar{1} \\ 0 & 0 & \bar{f} \\ 0 & 0 & 0 \end{pmatrix}$ | $(H+fK)\begin{pmatrix} 0 \\ 0 \\ 1 \end{pmatrix}$ |
| 46 | $M_{11}$ | $M_{41}$ | $(1\,f\,0)$ | $\begin{pmatrix} 0 & 0 & 0 \\ 0 & 0 & 0 \\ \bar{1} & \bar{f} & 0 \end{pmatrix}$ | $L\begin{pmatrix} 1 \\ f \\ 0 \end{pmatrix}$ |
| 47 | $M_{12}$ | $M_{42}$ | $(0\,0\,1)$ | $\begin{pmatrix} 0 & 0 & \bar{f} \\ 0 & 0 & \bar{1} \\ 0 & 0 & 0 \end{pmatrix}$ | $(fH+K)\begin{pmatrix} 0 \\ 0 \\ 1 \end{pmatrix}$ |
| 48 | $M_{12}$ | $M_{42}$ | $(f\,1\,0)$ | $\begin{pmatrix} 0 & 0 & 0 \\ 0 & 0 & 0 \\ \bar{f} & \bar{1} & 0 \end{pmatrix}$ | $L\begin{pmatrix} f \\ 1 \\ 0 \end{pmatrix}$ |
| 49 | $M_{21}$ | $M_{31}$ | $(0\,0\,1)$ | $\begin{pmatrix} 0 & 0 & 1 \\ 0 & 0 & \bar{f} \\ 0 & 0 & 0 \end{pmatrix}$ | $(\bar{H}+fK)\begin{pmatrix} 0 \\ 0 \\ 1 \end{pmatrix}$ |
| 50 | $M_{21}$ | $M_{31}$ | $(\bar{1}\,f\,0)$ | $\begin{pmatrix} 0 & 0 & 0 \\ 0 & 0 & 0 \\ 1 & \bar{f} & 0 \end{pmatrix}$ | $L\begin{pmatrix} \bar{1} \\ f \\ 0 \end{pmatrix}$ |
| 51 | $M_{22}$ | $M_{32}$ | $(0\,0\,1)$ | $\begin{pmatrix} 0 & 0 & f \\ 0 & 0 & \bar{1} \\ 0 & 0 & 0 \end{pmatrix}$ | $(f\bar{H}+K)\begin{pmatrix} 0 \\ 0 \\ 1 \end{pmatrix}$ |
| 52 | $M_{22}$ | $M_{32}$ | $(f\,\bar{1}\,0)$ | $\begin{pmatrix} 0 & 0 & 0 \\ 0 & 0 & 0 \\ f & \bar{1} & 0 \end{pmatrix}$ | $\bar{L}\begin{pmatrix} f \\ \bar{1} \\ 0 \end{pmatrix}$ |
| 53 | $M_{21}$ | $M_{41}$ | $(0\,1\,0)$ | $\begin{pmatrix} 0 & 1 & 0 \\ 0 & 0 & 0 \\ 0 & \bar{f} & 0 \end{pmatrix}$ | $(\bar{H}+fL)\begin{pmatrix} 0 \\ 1 \\ 0 \end{pmatrix}$ |
| 54 | $M_{21}$ | $M_{41}$ | $(\bar{1}\,0\,f)$ | $\begin{pmatrix} 0 & 0 & 0 \\ 1 & 0 & \bar{f} \\ 0 & 0 & 0 \end{pmatrix}$ | $K\begin{pmatrix} \bar{1} \\ 0 \\ f \end{pmatrix}$ |
| 55 | $M_{23}$ | $M_{43}$ | $(0\,1\,0)$ | $\begin{pmatrix} 0 & f & 0 \\ 0 & 0 & 0 \\ 0 & \bar{1} & 0 \end{pmatrix}$ | $(f\bar{H}+L)\begin{pmatrix} 0 \\ 1 \\ 0 \end{pmatrix}$ |
| 56 | $M_{23}$ | $M_{43}$ | $(f\,0\,\bar{1})$ | $\begin{pmatrix} 0 & 0 & 0 \\ f & 0 & \bar{1} \\ 0 & 0 & 0 \end{pmatrix}$ | $\bar{K}\begin{pmatrix} f \\ 0 \\ \bar{1} \end{pmatrix}$ |
| 57 | $M_{32}$ | $M_{42}$ | $(1\,0\,0)$ | $\begin{pmatrix} 0 & 0 & 0 \\ 1 & 0 & 0 \\ \bar{f} & 0 & 0 \end{pmatrix}$ | $(\bar{K}+fL)\begin{pmatrix} 1 \\ 0 \\ 0 \end{pmatrix}$ |
| 58 | $M_{32}$ | $M_{42}$ | $(0\,\bar{1}\,f)$ | $\begin{pmatrix} 0 & 1 & \bar{f} \\ 0 & 0 & 0 \\ 0 & 0 & 0 \end{pmatrix}$ | $H\begin{pmatrix} 0 \\ \bar{1} \\ f \end{pmatrix}$ |

| 59 | $M_{33}$ | $M_{43}$ | (1 0 0) | $\begin{pmatrix} 0 & 0 & 0 \\ f & 0 & 0 \\ \bar{1} & 0 & 0 \end{pmatrix}$ | $(f\bar{K}+L)\begin{pmatrix} 1 \\ 0 \\ 0 \end{pmatrix}$ |
| 60 | $M_{33}$ | $M_{43}$ | $(0\ f\ \bar{1})$ | $\begin{pmatrix} 0 & f & \bar{1} \\ 0 & 0 & 0 \\ 0 & 0 & 0 \end{pmatrix}$ | $\bar{H}\begin{pmatrix} 0 \\ f \\ \bar{1} \end{pmatrix}$ |

**Table 10** The same as Table 7 but for the case of PDWs, connecting domain pairs of the type *"R-semi-crossed"*.

| $N$ | $M_{m_1 n_1}$ | $M_{m_2 n_2}$ | $(hkl)$ | $[\Delta S]\frac{\Delta\alpha+\Delta\gamma}{2}$ | $[\Delta B]\frac{\Delta\alpha+\Delta\gamma}{2}$ |
|---|---|---|---|---|---|
| 61 | $M_{12}$ | $M_{23}$ | (0 1 1) | $\begin{pmatrix} 0 & \bar{2} & \bar{2} \\ 0 & \bar{p} & \bar{p} \\ 0 & p & p \end{pmatrix}$ | $(2H+pK+p\bar{L})\begin{pmatrix} 0 \\ 1 \\ 1 \end{pmatrix}$ |
| 62 | $M_{12}$ | $M_{23}$ | $(2\ p\ \bar{p})$ | $\begin{pmatrix} 0 & 0 & 0 \\ \bar{2} & \bar{p} & p \\ \bar{2} & \bar{p} & p \end{pmatrix}$ | $(K+L)\begin{pmatrix} 2 \\ p \\ \bar{p} \end{pmatrix}$ |
| 63 | $M_{13}$ | $M_{22}$ | (0 1 1) | $\begin{pmatrix} 0 & \bar{2} & \bar{2} \\ 0 & p & p \\ 0 & \bar{p} & \bar{p} \end{pmatrix}$ | $(2H+p\bar{K}+pL)\begin{pmatrix} 0 \\ 1 \\ 1 \end{pmatrix}$ |
| 64 | $M_{13}$ | $M_{22}$ | $(2\ \bar{p}\ p)$ | $\begin{pmatrix} 0 & 0 & 0 \\ \bar{2} & p & \bar{p} \\ \bar{2} & p & \bar{p} \end{pmatrix}$ | $(K+L)\begin{pmatrix} 2 \\ \bar{p} \\ p \end{pmatrix}$ |
| 65 | $M_{11}$ | $M_{33}$ | (1 0 1) | $\begin{pmatrix} \bar{p} & 0 & \bar{p} \\ \bar{2} & 0 & \bar{2} \\ p & 0 & p \end{pmatrix}$ | $(pH+2K+p\bar{L})\begin{pmatrix} 1 \\ 0 \\ 1 \end{pmatrix}$ |
| 66 | $M_{11}$ | $M_{33}$ | $(p\ 2\ \bar{p})$ | $\begin{pmatrix} \bar{p} & \bar{2} & p \\ 0 & 0 & 0 \\ \bar{p} & \bar{2} & p \end{pmatrix}$ | $(H+L)\begin{pmatrix} p \\ 2 \\ \bar{p} \end{pmatrix}$ |
| 67 | $M_{13}$ | $M_{33}$ | (1 0 1) | $\begin{pmatrix} p & 0 & p \\ \bar{2} & 0 & \bar{2} \\ \bar{p} & 0 & \bar{p} \end{pmatrix}$ | $(p\bar{H}+2K+pL)\begin{pmatrix} 1 \\ 0 \\ 1 \end{pmatrix}$ |
| 68 | $M_{13}$ | $M_{31}$ | $(\bar{p}\ 2\ p)$ | $\begin{pmatrix} p & \bar{2} & \bar{p} \\ 0 & 0 & 0 \\ p & \bar{2} & \bar{p} \end{pmatrix}$ | $(H+L)\begin{pmatrix} \bar{p} \\ 2 \\ p \end{pmatrix}$ |
| 69 | $M_{11}$ | $M_{42}$ | (1 1 0) | $\begin{pmatrix} \bar{p} & \bar{p} & 0 \\ p & p & 0 \\ \bar{2} & \bar{2} & 0 \end{pmatrix}$ | $(pH+p\bar{K}+2L)\begin{pmatrix} 1 \\ 1 \\ 0 \end{pmatrix}$ |
| 70 | $M_{11}$ | $M_{42}$ | $(p\,\bar{p}\,2)$ | $\begin{pmatrix} \bar{p} & p & \bar{2} \\ \bar{p} & p & \bar{2} \\ 0 & 0 & 0 \end{pmatrix}$ | $(H+K)\begin{pmatrix} p \\ \bar{p} \\ 2 \end{pmatrix}$ |
| 71 | $M_{12}$ | $M_{41}$ | $(1\,1\,0)$ | $\begin{pmatrix} p & p & 0 \\ \bar{p} & \bar{p} & 0 \\ \bar{2} & \bar{2} & 0 \end{pmatrix}$ | $(p\bar{H}+pK+2L)\begin{pmatrix} 1 \\ 1 \\ 0 \end{pmatrix}$ |
| 72 | $M_{12}$ | $M_{41}$ | $(\bar{p}\,p\,2)$ | $\begin{pmatrix} p & \bar{p} & \bar{2} \\ p & \bar{p} & \bar{2} \\ 0 & 0 & 0 \end{pmatrix}$ | $(H+K)\begin{pmatrix} \bar{p} \\ p \\ 2 \end{pmatrix}$ |
| 73 | $M_{21}$ | $M_{32}$ | $(\bar{1}\,1\,0)$ | $\begin{pmatrix} \bar{p} & p & 0 \\ \bar{p} & p & 0 \\ 2 & \bar{2} & 0 \end{pmatrix}$ | $(p\bar{H}+pK+2L)\begin{pmatrix} \bar{1} \\ 1 \\ 0 \end{pmatrix}$ |
| 74 | $M_{21}$ | $M_{32}$ | $(\bar{p}\,\bar{p}\,2)$ | $\begin{pmatrix} \bar{p} & \bar{p} & 2 \\ p & p & \bar{2} \\ 0 & 0 & 0 \end{pmatrix}$ | $(\bar{H}+K)\begin{pmatrix} \bar{p} \\ \bar{p} \\ 2 \end{pmatrix}$ |
| 75 | $M_{22}$ | $M_{31}$ | $(\bar{1}\,1\,0)$ | $\begin{pmatrix} p & \bar{p} & 0 \\ p & \bar{p} & 0 \\ 2 & \bar{2} & 0 \end{pmatrix}$ | $(pH+pK+2L)\begin{pmatrix} \bar{1} \\ 1 \\ 0 \end{pmatrix}$ |
| 76 | $M_{22}$ | $M_{31}$ | $(p\,p\,2)$ | $\begin{pmatrix} p & p & 2 \\ \bar{p} & \bar{p} & \bar{2} \\ 0 & 0 & 0 \end{pmatrix}$ | $(\bar{H}+K)\begin{pmatrix} p \\ p \\ 2 \end{pmatrix}$ |
| 77 | $M_{21}$ | $M_{43}$ | $(\bar{1}\,0\,1)$ | $\begin{pmatrix} \bar{p} & 0 & p \\ 2 & 0 & \bar{2} \\ \bar{p} & 0 & p \end{pmatrix}$ | $(p\bar{H}+2K+p\bar{L})\begin{pmatrix} \bar{1} \\ 0 \\ 1 \end{pmatrix}$ |
| 78 | $M_{21}$ | $M_{43}$ | $(\bar{p}\,2\,\bar{p})$ | $\begin{pmatrix} \bar{p} & 2 & \bar{p} \\ 0 & 0 & 0 \\ p & \bar{2} & p \end{pmatrix}$ | $(\bar{H}+L)\begin{pmatrix} \bar{p} \\ 2 \\ \bar{p} \end{pmatrix}$ |
| 79 | $M_{23}$ | $M_{41}$ | $(\bar{1}\,0\,1)$ | $\begin{pmatrix} p & 0 & \bar{p} \\ 2 & 0 & \bar{2} \\ p & 0 & \bar{p} \end{pmatrix}$ | $(pH+2K+pL)\begin{pmatrix} \bar{1} \\ 0 \\ 1 \end{pmatrix}$ |
| 80 | $M_{23}$ | $M_{41}$ | $(p\,2\,p)$ | $\begin{pmatrix} p & 2 & p \\ 0 & 0 & 0 \\ \bar{p} & \bar{2} & \bar{p} \end{pmatrix}$ | $(\bar{H}+L)\begin{pmatrix} p \\ 2 \\ p \end{pmatrix}$ |
| 81 | $M_{32}$ | $M_{43}$ | $(0\,\bar{1}\,1)$ | $\begin{pmatrix} 0 & 2 & \bar{2} \\ 0 & \bar{p} & p \\ 0 & \bar{p} & p \end{pmatrix}$ | $(2H+p\bar{K}+p\bar{L})\begin{pmatrix} 0 \\ \bar{1} \\ 1 \end{pmatrix}$ |
| 82 | $M_{32}$ | $M_{43}$ | $(2\,\bar{p}\,\bar{p})$ | $\begin{pmatrix} 0 & 0 & 0 \\ 2 & \bar{p} & \bar{p} \\ \bar{2} & p & p \end{pmatrix}$ | $(L+\bar{K})\begin{pmatrix} 2 \\ \bar{p} \\ \bar{p} \end{pmatrix}$ |
| 83 | $M_{33}$ | $M_{42}$ | $(0\,\bar{1}\,1)$ | $\begin{pmatrix} 0 & 2 & \bar{2} \\ 0 & p & \bar{p} \\ 0 & p & \bar{p} \end{pmatrix}$ | $(2H+K+L)\begin{pmatrix} 0 \\ \bar{1} \\ 1 \end{pmatrix}$ |

| 84 | $M_{33}$ | $M_{42}$ | $(2\,p\,p)$ | $\begin{pmatrix} 0 & 0 & 0 \\ 2 & p & p \\ \bar{2} & \bar{p} & \bar{p} \end{pmatrix}$ | $(L+\bar{K})\begin{pmatrix} 2 \\ p \\ p \end{pmatrix}$ |
|---|---|---|---|---|---|

Table 7, 8, 9, 10 reveal that certain W-Walls have the same orientations. For instance, all domain pairs type "*R-planar*" $M_{11}M_{21}, M_{31}M_{41}$ and all domain pairs of the type "*R-semi-planar*", $M_{12}, M_{22}, M_{13}, M_{23}$ have (100)-oriented PDWs. Table 11 presents all the distinct PDW orientations and their relevant details. It reveals that all the PDWs belong to five orientation families $\{100\}, \{110\}, \{2rr\}, \{10f\}, \{2pp\}$, so that PDWs of 45 distinct orientations are present. Furthermore, the table demonstrates the distribution of PDWs based on the pair type and the angle between the polarization direction. It indicates that 84 PDWs are classified into 12 DWs, 30 DWs, 30 DWs, 12DWs with the angles between spontaneous polarization direction close to 0, 71, 109 and 180°, respectively.

**Table 11** The orientation families of PDWs and their distribution between DW of different types. The first column contains the identifier of the family where {} indicate the list of $m3m$ equivalent orientations e.g. $\{110\}$ means the list of $(011), (101), (110), (01\bar{1}), (10\bar{1})$ and $(01\bar{1})$. The second column contains the number of different orientations. The third column contains the number of PDWs of the specific orientation family. The remaining columns show the distribution of these PDWs according to the pair type and the "zero-charge" angle between polarization directions.

| $\{hkl\}$ | $M$ | N Walls | $N_0$ | $N_{RP\ 71}$ | $N_{RSP\ 71}$ | $N_{RSC\ 71}$ | $N_{RP\ 109}$ | $N_{RSP\ 109}$ | $N_{RSC\ 109}$ | $N_{180}$ |
|---|---|---|---|---|---|---|---|---|---|---|
| $\{100\}$ | 3 | 18 | – | – | – | – | 6 | 12 | – | – |
| $\{110\}$ | 6 | 30 | – | 6 | – | 12 | – | – | – | 12 |
| $\{2rr\}$ | 12 | 12 | 12 | – | – | – | – | – | – | – |
| $\{f01\}$ | 12 | 12 | – | – | 12 | – | – | – | – | – |
| $\{2pp\}$ | 12 | 12 | – | – | – | – | – | – | 12 | – |
| All walls | 45 | 84 | 12 | | 30 | | | 30 | | 12 |

Figure 7 displays the orientation of all the PDWs for different choices of lattice parameters. The normal vectors to these walls are shown using the poles on the stereographic projection. The W-walls are marked by the poles with a solid line edge and the color of the pole reflects the angle between the spontaneous polarization directions being close to 0, 71, 109 and 180° (as specified in the last column of Table 5). Each stereographic projection, from left to right, shows DWs between the domain pairs of the type's "*R-sibling*", "*R-planar*", "*R-semi-planar*" and "*R-semi-crossed*". The supporting material includes the animated version of this film showing how the orientation of these DWs changes with the lattice parameters.

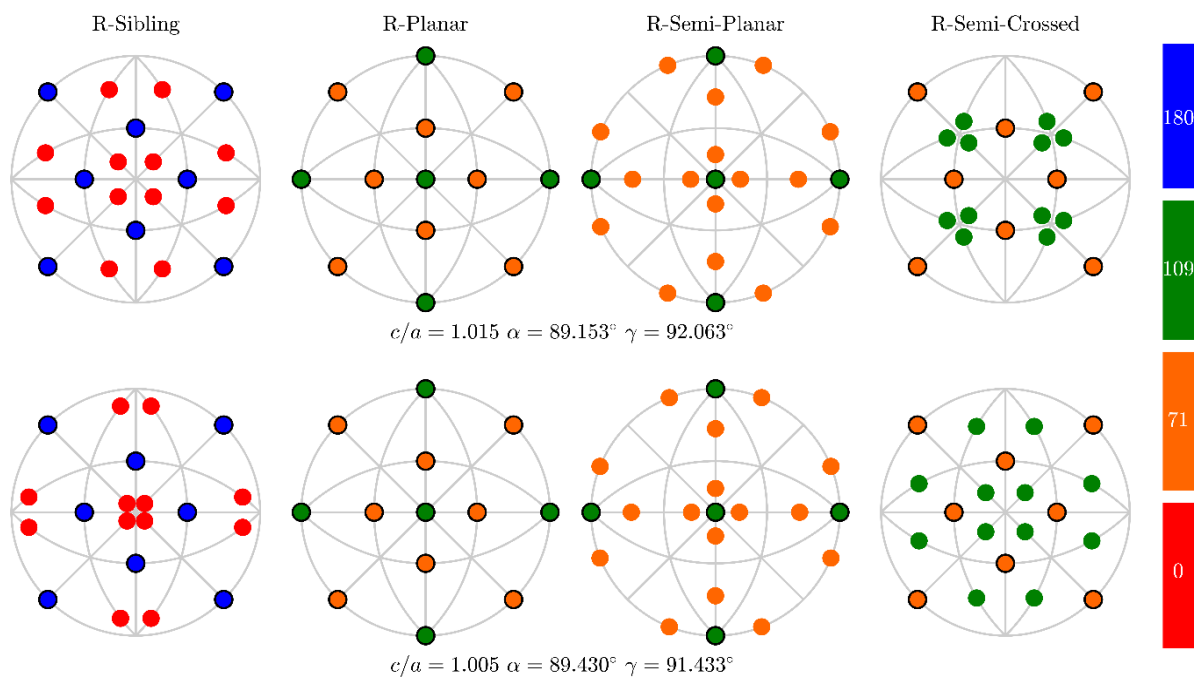

**Figure 7** The orientation of the DW is normal for all the PDWs, which connect domain pairs of the "*R-sibling*", "*R-planar*", "*R-semi-planar*" and "*R-semi-crossed*" types, with 45 different orientations in total. These orientations are distributed among five different orientation families. The normals are shown using the poles on the stereographic projection viewed along the direction [001] with the poles corresponding to the W-walls framed by solid line. The lattice parameters are chosen arbitrarily. The supporting information includes animated versions of the same figure for different values of the monoclinic lattice parameters.

## 6. Conclusion

We have applied the theory of PDWs to create a list of 84 PDWs connecting ferroelastic domains of monoclinic ($Cm/Cc$) symmetry. Our list includes analytical expressions for the Miller indices of the PDWs, matrices of transformation between the corresponding pseudocubic basis vectors and expressions for the reciprocal space separation between the corresponding Bragg peak pairs. The 84 PDWs can have 45 different orientations and are grouped into five orientational families.

Our derivation of this list assumed that the two-step transition from the cubic ($Pm\bar{3}m$) phase to the monoclinic ($Cm/Cc$) phase results in the formation of 12 ferroelastic monoclinic domains. The first step of this transition (from the cubic to the rhombohedral $R3m/R3c$ phase) results in the formation of four ferroelastic domains, while the second step (from the rhombohedral to the monoclinic phase) splits each of them into groups of three monoclinic domains. We identified five different types of domain pairs (referred here as "R-sibling", "R-planar", "R-semi-planar", "R-semi-crossed" and "R-crossed") each with its own expression for the PDW orientation. As shown in the previous works (Fousek & Janovec, 1969; Sapriel, 1975), we found that the crystallographic orientation / Miller indices of PDWs can be fixed (for the so-called *W*-walls) or depend on the values of the monoclinic lattice parameters (for the so-called *S*-walls). We found that the orientation of such walls can be controlled by the three simple parameters $r = \frac{2}{\Delta\alpha - \Delta\gamma}\left(\frac{c}{a} - 1\right)$, $f = \frac{\Delta\gamma}{\Delta\alpha}$ and $p = \frac{2}{\Delta\alpha + \Delta\gamma}\left(\frac{c}{a} - 1\right)$.

We have demonstrated that the rotatable domain walls can be described by the "Miller indices" $\{2rr\}, \{10f\}, \{2p\bar{p}\}$. Even small change in the monoclinic distortion (such as $\frac{c}{a}, \Delta\alpha, \Delta\gamma$) can cause a significant rotation of the PDW. This process is often referred to as "thermal switching". Furthermore, we have predicted the angles between polarization directions for the cases when DWs are not charged.

The results of this work can be useful in several different ways. First, the availability of simple analytical expressions (Tables 7-10) for the orientation of domain walls can help in describing the domain switching through DW rotation or DW motion. Such a process can be induced by the change of the temperature or external electric field, for example. Second, the expressions for the separation between Bragg peak (Tables 7 – 10) can help investigate monoclinic domain patterns, using "single crystal" X-ray diffraction. Third, the expressions may be useful for the precise calculation of the angles between SPDs of various domains. Such angles can be easily evaluated using the corresponding matrices of transformation between the domains basis vectors in Tables 7 – 10.

The results hold significant importance in the analysis of domains within crystals and epitaxial thin films. Indeed, the observation of monoclinic domains in epitaxial thin film is common (see e.g. Schmidbauer *et al.*, 2017; Gaal *et al.*, 2023) where one or another type of monoclinic distortion is stabilized by the substrate-film lattice mismatch. Modulating this mismatch can influence the monoclinic lattice parameters and consequently, the orientation of PDWs between them. It is worth highlighting that certain distinctions may arise due to variations in the number of monoclinic domains present. In the case of "free-standing" single crystals, the phase transition sequence from cubic to rhombohedral to monoclinic ideally results in the presence of 12 equivalent domains. However, introducing bias at any of these transitional stages can alter this configuration. For instance, the application of an electric field along the pseudocubic [111] direction during the cubic-to-rhombohedral phase transition may lead to the formation of just one rhombohedral domain instead of the expected four. Subsequently, the rhombohedral-to-monoclinic transition further divides this domain into three monoclinic domains. Consequently, in such scenarios, only "R-sibling" domain pairs, connected by six PDWs, must be considered. The presence of the substrate can bias or suppress the formation of specific domains, such as favoring the presence of domain pairs of the R-sibling type exclusively, and this, in turn, can impact the number of PDWs. A detailed characterization of PDWs in relation to the origin of these domains can prove useful for cataloging the potential PDWs existing between thin film domains or in other cases when formation of domains is biased or engineered.

Finally, this article describes the PDWs between monoclinic domains of $M_A/M_B$ type. A similar formalism for the monoclinic $M_C$ symmetry case will be presented in the follow-up publication.

## Appendix A. The list of notations and most important crystallographic relationships

This paper uses the notations from Gorfman *et al.* (2022). For the convenience of the reader the most important of them are also summarized here.

<u>Basis vectors:</u> $\boldsymbol{a}_{im}$ ($i = 1\ldots3$) are the basis vectors of a crystal lattice. The second index refers to the ferroelastic domain variant $m$. $m = 0$ corresponds to the crystal lattice of higher symmetry (e.g., cubic) "parent" phase (Figure 1). The parallelepiped based on the vectors $\boldsymbol{a}_{im}$ forms a unit cell.

<u>Unit-cell settings:</u> Many unit-cell settings exist for the same lattice (Gorfman, 2020). Here, we prefer the cell settings $\boldsymbol{a}_{im}$ ($m > 0$) obtained by the smallest possible distortion / rotation of the parent-phase basis vectors $\boldsymbol{a}_{i0}$.

<u>Metric tensor / matrix of dot products:</u> $G_{ij} = \boldsymbol{a}_i\boldsymbol{a}_j$ is the metric tensor (Giacovazzo, 1992; Hahn, 2005). The corresponding $3 \times 3$ matrix $[G]_m$ is the matrix of dot products for the domain variant $m$. Their determinants are $|G| = V_A^2$ ($V_A$ is the unit cell volume).

<u>The transformation matrix:</u> the transformation e.g., from the basis vectors $\boldsymbol{a}_{im}$ to the basis vectors $\boldsymbol{a}_{in}$ is defined by the $3 \times 3$ transformation matrix $[S]$. The columns of the matrix $[S]$ are the coordinates of $\boldsymbol{a}_{in}$ with respect to $\boldsymbol{a}_{im}$.

$$(\boldsymbol{a}_{1n}\quad \boldsymbol{a}_{2n}\quad \boldsymbol{a}_{3n}) = (\boldsymbol{a}_{1m}\quad \boldsymbol{a}_{2m}\quad \boldsymbol{a}_{3m})\begin{pmatrix}S_{11} & S_{12} & S_{13}\\ S_{21} & S_{22} & S_{23}\\ S_{31} & S_{32} & S_{33}\end{pmatrix} \qquad (A1)$$

<u>Transformation of the metric tensor:</u> the transformation of the basis vectors (A1) leads to the following transformation of the corresponding metric tensors.

$$[G]_n = [S]^T[G]_m[S] \qquad (A2)$$

This relationship can be extended to any cases of transformation between coordinate systems.

<u>The difference transformation matrix</u> is defined as the difference between $[S]$ and the unitary matrix $[I]$

$$[\Delta S] = [S] - [I] \qquad (A3)$$

<u>Twinning matrix:</u> $[T]$ represents a symmetry operation of the parent phase lattice (i.e., the one built using the basis vectors $\boldsymbol{a}_{i0}$) that is no longer the symmetry operation of a ferroelastic phase lattice. We define $[T]$ as $3 \times 3$ matrix, which describes the transformation to the coordinate system $\boldsymbol{a}_{i0}$ from its symmetry equivalent $\boldsymbol{a}'_{i0}$ using the following formal matrix equation:

$$(\boldsymbol{a}_{10}\quad \boldsymbol{a}_{20}\quad \boldsymbol{a}_{30}) = (\boldsymbol{a}'_{10}\quad \boldsymbol{a}'_{20}\quad \boldsymbol{a}'_{30})\begin{pmatrix}T_{11} & T_{12} & T_{13}\\ T_{21} & T_{22} & T_{23}\\ T_{31} & T_{32} & T_{33}\end{pmatrix}, \qquad (A4)$$

The number of symmetry equivalent coordinate systems is equal to the order of the holohedry point symmetry group (e.g., 48 for a cubic lattice). The transition from a paraelastic to a ferroelastic phase is associated with the distortion of the basis vectors $\boldsymbol{a}_{i0} \longrightarrow \boldsymbol{a}_{im}$. Such distortion, however, can commence from any of the symmetry equivalent $\boldsymbol{a}'_{i0}$. Let us assume that $\boldsymbol{a}_{i0}$ and $\boldsymbol{a}'_{i0}$ serve as the starting points for domain variants $m$ and $n$ correspondingly. The following relationship between $[G_n]$ and $[G_m]$ exists:

$$[G]_n = [T]^T [G]_m [T] \tag{A5}$$

Reciprocal basis vectors: The superscript * refers to the reciprocal bases, e.g., $\boldsymbol{a}_i^*$ are such that $\boldsymbol{a}_i \boldsymbol{a}_j^* = \delta_{ij}$ The reciprocal metric tensor is $G_{ij}^* = \boldsymbol{a}_i^* \boldsymbol{a}_j^*$. The relationship $[G^*] = [G]^{-1}$ holds.

Transformation between the reciprocal basis vectors: If the direct basis vectors (e.g., $\boldsymbol{a}_{im}$ and $\boldsymbol{a}_{in}$) are related by the matrix $[S]$ (according to equation (A1)) then the corresponding reciprocal lattice vectors $\boldsymbol{a}_{im}^*$ and $\boldsymbol{a}_{in}^*$) are related by the matrix $[S^*]$. The following relationship between $[S]$ and $[S^*]$ holds:

$$[S^*]^T = [S]^{-1} \tag{A6}$$

The difference transformation matrix between the reciprocal basis vectors is defined according to the equation:

$$[\Delta S^*] = [S^*] - [I] \tag{A7}$$

# Supporting information

Supporting information contains the animated version of Figure 7, which shows how the orientation of domain walls changes with the lattice parameters.